\documentclass[a4paper]{article}
\usepackage[a4paper, total={6.2in, 9in}]{geometry}
\usepackage{floatrow}
\usepackage{hyperref}
\usepackage{filecontents}
\usepackage{subfigure}
\usepackage{bm}
\usepackage[font={scriptsize,bf}]{caption}
\usepackage{amsmath}
\usepackage{amsfonts}
\usepackage{amssymb}
\usepackage{graphicx}
\usepackage{bm}
\usepackage{multicol}
\usepackage{multirow}
\usepackage{epstopdf}
\usepackage{soul}
\usepackage{pifont}
\usepackage{lineno}
\usepackage[table]{xcolor}
\usepackage{color}
\usepackage[autostyle]{csquotes}
\usepackage{tikz,pgf}
\usepackage{collcell}

\usepackage{pgfplots}
\usepackage{wasysym}
\usepackage{lipsum}
\usepackage{algorithm} 
\usepackage{algpseudocode} 
\newlength\fheight
\newlength\fwidth
\newcommand{\cmmnt}[1]{\ignorespaces}
\usepackage{etoolbox}
\usepackage{setspace}
\usepackage{csquotes}
\usepackage{cancel}
\DeclareMathOperator{\sgn}{sgn}

\title{Phase Reduction Analysis of Transonic Buffet}
\author{ \large Michael Candon$^{1,2}$\thanks{corresponding author, candon.michael@rmit.edu.au} and Earl Dowell$^1$}

\date{
	\normalsize 
    $^1$Duke University, Durham, NC, 27708\\
    $^2$Department of Aerospace Engineering, RMIT University, Melbourne, AUS, 3000\\[2ex]%
}

\begin{document}
	
	\maketitle
	\begin{abstract}

        This paper develops a phase-reduced description of transonic buffet, treating buffet as a stable autonomous fluid limit cycle that is weakly perturbed by harmonic structural motion. When combined with aerodynamic work, the phase reduction model yields closed-form expressions for the equivalent aerodynamic damping and stiffness, together with a unified description of phase locking, phase slipping and aeroelastic stability of the fluid-structural system. The resulting analysis provides new insight into the mechanisms governing aeroelastic instability in oscillatory flows, including aerodynamic damping maps, singular behaviour at zero detuning, and broad regions of subcritical aeroelastic instability.
    \end{abstract}

\section{Introduction}
\label{sec:intro}
Self-excited oscillatory fluid flows can interact strongly with structural motion, producing complex nonlinear fluid--structure interactions. In such systems, the response is governed by the natural frequencies of the structure, the natural frequencies of the buffeting flow, and also by their evolving relative phase, which determines the exchange of energy between the fluid and the structure. While it is well known that this interaction can produce aeroelastic instabilities, synchronisation, and limit cycle oscillations, a direct connection between synchronisation dynamics and conventional aeroelastic quantities such as equivalent aerodynamic damping and stiffness has not been previously been demonstrated.

Transonic buffet is a canonical and reasonably complex example of a self-excited fluid instability, characterised by large self-sustained shock oscillations and unsteady separation and re-attachment of the boundary layer~\cite{tijdeman77, lee01, crouch09}. Buffet represents an important operating-envelope limitation for commercial aircraft, and a performance, fatigue and handling concern for military aircraft~\cite{levinski20}. Although buffet on rigid aerofoils and wings has been studied extensively, its interaction with structural motion is less well understood~\cite{giannelis17}. Buffet--structure interactions are commonly characterised by the \enquote{frequency ratio} $\hat{f}_n=\omega_n/\omega_B$, where $\omega_n$ is the imposed frequency or the natural frequency of structural mode $n$, and $\omega_B$ is the buffet frequency.

Forced-motion studies have demonstrated synchronisation of the buffet oscillation with structural motion both above and below $\hat{f}_n=1$~\cite{raveh09,hartmann13}. Numerical studies of elastically suspended aerofoils have subsequently shown that instability and lock-in may occur in both single-~\cite{quan15,giannelis16,gao17} and multi-degree-of-freedom~\cite{raveh14,candon25f} systems when the frequency ratio is near unity. Within the lock-in region, the fluid and structural oscillations synchronise at the structural natural frequency and the structural response experiences very large amplification. Outside this region, the response is generally much smaller and oscillates at the buffet frequency. Whether aeroelastic instability and synchronisation occur above or below $\hat{f}_n=1$ depends on the structural mode shape and the associated generalised aerodynamic forcing~\cite{candon26b}. Gao \textit{et al.} \cite{gao17} interpret this aeroelastic instability as a form of single-degree-of-freedom flutter~\cite{dowell24}, which is supported by Candon \textit{et al.}\cite{candon25f} who demonstrate that the associated aerodynamic damping remains negative down to the small-amplitude limit. However, the role of synchronisation dynamics in this limit remains unclear, and a direct relationship with aerodynamic work and the resulting aeroelastic behaviour has not been formally established.

Treating the problem as a weak interaction between an autonomous fluid oscillator and harmonic structural motion, phase reduction provides a useful framework for the analysis. Near a stable limit cycle, the high-dimensional fluid dynamics can be reduced to a single phase variable whose evolution under structural perturbations is governed by a phase sensitivity function (PSF). While phase reduction has been widely used in electrical engineering, biology, and nonlinear dynamics for many years~\cite{adler46, winfree67, kuramoto84, nakao16}, it has only more recently been applied in fluid dynamics, largely through the work of Kawamura, Taira, and their collaborators~\cite{kawamura15,taira18,nair21,kawamura22,taira26}. In these studies, the PSF may be obtained either from direct impulsive perturbations of the fluid limit cycle~\cite{kawamura15,taira18,nair21} or through adjoint methods~\cite{kawamura22}. The application of phase reduction to compressible self-excited flows, however, remains unexplored.

The present work applies phase reduction to transonic buffet and develops a mathematical connection between the phase-reduced dynamics and aerodynamic work. The phase sensitivity of the buffet limit cycle is identified from a small number of impulsive perturbations applied through the structural coordinates. Assuming harmonic structural motion, the resulting phase coupling function (PCF) is combined with the aerodynamic work relation to derive closed-form frequency-domain expressions for the equivalent aerodynamic damping and stiffness.

The formulation is assessed using two-dimensional unsteady Reynolds-averaged Navier--Stokes simulations of buffet over the ONERA OAT15A aerofoil. Phase sensitivities are identified for both heave and pitch perturbations, and the phase reduced model prediction is compared with prescribed-motion and coupled aeroelastic simulations. The analysis describes phase locking and slipping, stability boundaries, stable and unstable limit cycle branches, and shows how the relative phase between the buffet oscillation and structural motion governs the aerodynamic energy transfer. The phase-reduced aerodynamic damping and stiffness equations naturally produce singular aerodynamic damping at zero detuning, and subcritical aeroelastic instabilities, connecting the linear and nonlinear flutter dynamic pressure and amplitude to conventional structural and aerodynamic parameters.

\section{Phase Reduction and Closed-Form Aerodynamic Damping and Stiffness}
\label{sec:derivation}

\subsection{Definitions and Reference Phase}

The variables $q$ and $Q$ are used to denote the imposed generalised structural coordinate and its corresponding generalised aerodynamic force, respectively. The imposed harmonic motion is described by an amplitude $A$, angular frequency $\Omega$, and phase $\phi=\Omega t$. The unforced buffet oscillation is characterised by an angular frequency $\omega_b$, a reference phase $\theta_b=\omega_b t$, and a first-harmonic aerodynamic force amplitude $\hat{Q}_B$. The harmonic motion is written as

\begin{equation}
\label{eq:1}
q(t)=A\sin\phi, \qquad \dot{q}(t)=A\Omega\cos\phi
\end{equation}

Assuming that the forced buffet dynamics remain close to its autonomous periodic orbit and that the aerodynamic response is dominated by its first harmonic, the total oscillatory aerodynamic response (including both the autonomous and imposed motion contributions) may be approximated as

\begin{equation}
\label{eq:2}
Q_1(t)=\hat{Q}\cos\left(\theta(t)+\chi_Q\right)
\end{equation}

where $\hat{Q}$ is the amplitude of the first harmonic of the aerodynamic response and $\theta(t)$ is the instantaneous buffet phase, the origin of which is arbitrary. For the unforced buffet, $\theta=\theta_b$, whereas when forced its phase evolves and $\dot{\theta}$ can depart from $\omega_b$. The quantity $\chi_Q$ is the phase offset of the aerodynamic force first-harmonic relative to this buffet phase coordinate.

\subsection{Phase Reduction and the Adler Equation}
\label{sec:phasereduction}

Let the full fluid state be $\mathbf{x}(t)$ and let the motion-induced perturbation entering the fluid equations be $\mathbf{f}(t;A,\Omega)$. A general forced fluid model may be written as

\begin{equation}
\label{eq:3}
\dot{\mathbf{x}}=\mathbf{F}(\mathbf{x})+\mathbf{B}(\mathbf{x})\mathbf{f}(t)
\end{equation}

where $\mathbf{F}()$ is the autonomous fluid vector field and $\mathbf{B}()$ maps the structural input into the fluid equations. Assuming that the unforced fluid system possesses a stable periodic buffet orbit, $\mathbf{x}_0(\theta)$, parameterised by the phase $\theta$ and angular frequency $\omega_b$, first-order phase reduction under weak perturbation gives

\begin{equation}
\label{eq:4}
\dot{\theta}=\omega_b+\mathbf{Z}(\theta)\mathbin{\cdot}\mathbf{f}(t)
\end{equation}

\noindent where $\mathbf{Z}(\theta)$ is the phase-sensitivity function, or phase-response function. Introducing the buffet--aerofoil relative motion phase as $\psi(t)=\theta(t)-\phi$, and expressing the forcing in terms of its phase, gives

\begin{equation}
\label{eq:5}
\dot{\psi}(t)=\omega_b-\Omega+\mathbf{Z}\bigl(\phi+\psi(t)\bigr)\mathbin{\cdot}\mathbf{f}(\phi)
\end{equation}

Under weak forcing and near resonance, $\psi(t)$ evolves slowly relative to the forcing phase $\phi$. Averaging over one forcing cycle gives the phase-difference model

\begin{equation}
\label{eq:6}
\dot{\psi}(t)=\Delta\omega+\Gamma\bigl(\psi(t)\bigr)
\end{equation}

\noindent where the detuning is $\Delta\omega=\omega_b-\Omega$ and the phase-coupling function is

\begin{equation}
\label{eq:7}
\Gamma(\psi)=\frac{1}{2\pi}\int_0^{2\pi}\mathbf{Z}(\phi+\psi)\mathbin{\cdot}\mathbf{f}(\phi)\,d\phi
\end{equation}

\noindent Because $\Gamma(\psi)$ is $2\pi$-periodic, retaining the mean and first harmonic of its Fourier expansion gives

\begin{equation}
\label{eq:8}
\Gamma(\psi) \approx \Gamma^{(1)}(\psi) =\Gamma_0-K\sin(\psi+\beta)
\end{equation}

\noindent \noindent where $K\geq0$ and $\beta$ are the amplitude and phase of the first harmonic, respectively. The first-order phase reduction used to obtain $K$ neglects nonlinear forcing-amplitude effects, introducing an error of $\mathcal{O}(A^2)$ relative to the full forced dynamics. The mean term $\Gamma_0$ represents the phase-independent frequency shift induced by the perturbation, while the sinusoidal term represents its phase-dependent contribution. Substituting Eq.~\ref{eq:8} into Eq.~\ref{eq:6}, defining the effective detuning as $\Delta=\Delta\omega+\Gamma_0$, and using $\varphi(t)=\psi(t)+\beta$ gives the Adler equation~\cite{adler46}:

\begin{equation}
\label{eq:9}
\dot{\varphi}(t)=\Delta-K\sin\varphi(t)
\end{equation}

The aerodynamic forces can be expressed in consistent phase coordinates via substitution of the Adler phase coordinate $\theta(t) = \varphi(t) + \phi - \beta$ into Eq.~\ref{eq:2} giving

\begin{equation}
\label{eq:10}
Q_1(t)=\hat{Q}\cos\left(\phi+\varphi(t)+\chi\right), \qquad \chi=\chi_Q-\beta
\end{equation}

\subsection{Equivalent Aerodynamic Damping from Aerodynamic Work}

Starting by equating the work done by an equivalent viscous aerodynamic force, $W_{\mathrm{viscous}}$, to the work done by the actual aerodynamic force, $W_{\mathrm{aero}}$

\begin{equation}
\label{eq:11}
W_{\mathrm{viscous}} = W_{\mathrm{aero}}
\end{equation}

\noindent and using a long-time average, the aerodynamic work-per-cycle from  Eqs.~\ref{eq:1} and \ref{eq:10} is given as

\begin{equation}
\label{eq:12}
\begin{aligned}
W_{\mathrm{aero}}
&=\frac{2\pi}{\Omega}\lim_{T\rightarrow\infty}\frac{1}{T}\int_0^T Q_1(t)\dot{q}(t)\,dt \\
&=2\pi A\hat{Q}\lim_{T\rightarrow\infty}\frac{1}{T}\int_0^T\cos\phi\cos\left(\phi+\varphi(t)+\chi\right)\,dt \\
&=\pi A \hat{Q}\left[\lim_{T\rightarrow\infty}\frac{1}{T}\int_0^T\cos\left(\varphi(t)+\chi\right)\,dt+\cancelto{0}{\lim_{T\rightarrow\infty}\frac{1}{T}\int_0^T\cos\left(2\phi+\varphi(t)+\chi\right)\,dt}\right]
\end{aligned}
\end{equation}

\noindent where $T$ is the total long-time averaging window. Then under the weak-coupling and near-resonance assumptions, $\varphi(t)$ evolves slowly relative to $\phi$ and the second term in Eq.~\ref{eq:12} averages to zero over many forcing periods. Using an equivalent viscous force, $Q_{\mathrm{viscous}}=-c_a\dot{q}$, the viscous work is defined as

\begin{equation}
\label{eq:13}
W_{\mathrm{viscous}} = \frac{2\pi}{\Omega}\lim_{T\rightarrow\infty}\frac{1}{T}\int_0^T c_a \dot{q}(t)^2 dt \approx  c_a \pi A^2\Omega
\end{equation}

\noindent and substituting Eqs.~\ref{eq:12} and Eq.~\ref{eq:13} into Eq.~\ref{eq:11}, the equivalent aerodynamic damping coefficient becomes

\begin{equation}
\label{eq:14}
c_a=-\frac{\hat{Q}}{A\Omega} \lim_{T\rightarrow\infty} \frac{1}{T} \int_0^T\cos\left(\varphi(t)+\chi\right)\,dt
\end{equation}

\noindent which is the starting point for the phase-locked and phase-slipping damping expressions.

\subsection{Phase-Locked Damping}

A locked state means that the buffet-aerofoil relative phase difference is constant, $\dot \varphi (t) = 0$, $\varphi (t) = \varphi^*$, and Eq.~\ref{eq:14} reduces to

\begin{equation}
\label{eq:15}
 c_a^{\mathrm{lock}} = -\frac{\widehat Q}{A\Omega}\cos(\varphi^*+\chi).
\end{equation}

\noindent From the phase reduction perspective, a phase-locked state also corresponds to a fixed point of Eq.~\ref{eq:9}, such that

\begin{equation}
\label{eq:16}
\sin\varphi^*=\frac{\Delta}{K}, \qquad \cos\varphi^* = \sqrt{1-\left(\frac{\Delta}{K}\right)^2}, \qquad |\Delta|\leq K
\end{equation}

\noindent where $|\Delta|\leq K$ is known as the \enquote{Adler condition} for phase locking. Physically, $|\Delta|\leq K$ is stating that lock-in will occur if the phase dependent frequency shift induced by the perturbation, $K$, overcomes the detuning frequency, $\Delta$. Then using $\cos(\varphi^*+\chi) = \cos\varphi^*\cos\chi - \sin\varphi^*\sin\chi$, and substituting into Eq.~\ref{eq:15} one can obtain

\begin{equation}
\label{eq:17}
c_a^{\mathrm{lock}}= -\frac{\widehat Q}{A\Omega} \left[\sqrt{1-\left(\frac{\Delta}{K}\right)^2} \cos\chi-\frac{\Delta}{K}\sin\chi\right], \quad |\Delta|\leq K
\end{equation}

This description has several useful properties. Since \(K\propto A\), this implies that for any fixed nonzero detuning, lock-in is lost below a finite forcing amplitude. \textbf{Therefore, a locked damping expression cannot be continued to \(A\to 0\) when \(\Delta\omega\neq 0\), and a damping singularity it not possible.} By contrast, when \(\Delta\omega=0\), the lock-in condition is automatically satisfied and Eq.~\ref{eq:17} becomes

\begin{equation}
\label{eq:18}
	c_a^{\mathrm{lock}}(\Omega=\omega_b) =-\frac{\hat Q}{A\omega_b}\cos\chi
\end{equation}

\textbf{Therefore, given that \(\hat Q\to \hat{Q}_{B}\) as \(A\to 0\), and assuming that \(\cos\chi\neq 0\), the equivalent aerodynamic damping has a \(1/A\) singularity at zero detuning.}

\subsection{Phase-Slipping Damping}

Now the damping equation for the phase slipping case is derived, which in practice is more important than the phase locked case. The reason for this is that the buffet induced aeroelastic instabilities are initially driven by negative aerodynamic damping values that occur when the structural response amplitude is very small and before lock-in has occurred. 

In terms of the Adler equation, slipping occurs when $|\Delta|>K$. In this regime, $\varphi(t)$ advances or retreats by $2\pi$ during each complete phase-slip cycle, which generally spans many structural cycles. From Eq.~\ref{eq:9}, the duration of one complete slip, $T_\mathrm{slip}$, is

\begin{equation}
\label{eq:19}
dt = \frac{d\varphi}{|\Delta-K\sin\varphi|} \implies T_{\mathrm{slip}} = \int_0^{2\pi} \frac{d\varphi}{|\Delta-K\sin\varphi|} = \frac{2\pi}{\sqrt{\Delta^2-K^2}}
\end{equation}

The phase slipping damping equation leverages the periodicity of the phase-slipping dynamics and Eq.~\ref{eq:14} is evaluated over one complete slip

\begin{equation}
\label{eq:20}
 c_a^{\mathrm{slip}} = -\frac{\hat Q}{A\Omega}\frac{1}{T_{\mathrm{slip}}}\int_0^{T_{\mathrm{slip}}}\cos\bigl(\varphi(t)+\chi\bigr)dt
\end{equation}

\noindent Changing the integration variable from $t$ to $\varphi$, substituting Eq.~\ref{eq:19} into Eq.~\ref{eq:20} gives 

\begin{equation}
\label{eq:21}
 c_a^{\mathrm{slip}} = -\frac{\hat Q}{A\Omega}\frac{\displaystyle\int_0^{2\pi}\frac{\cos(\varphi+\chi)}{|\Delta-K\sin\varphi|}\,d\varphi}{\displaystyle\int_0^{2\pi}\frac{1}{|\Delta-K\sin\varphi|}\,d\varphi}
\end{equation}

\noindent Expanding the numerator of Eq.~\ref{eq:21} using $\cos(\varphi+\chi) = \cos\varphi\cos\chi - \sin\varphi\sin\chi$ gives

\begin{equation} 
\label{eq:22} 
c_a^{\mathrm{slip}}=-\frac{\hat Q}{A\Omega}\frac{I_c\cos\chi-I_s\sin\chi}{I_0} 
\end{equation}

\noindent where

\begin{equation}
\label{eq:23}
I_0 = \int_0^{2\pi} \frac{d\varphi}{|\Delta-K\sin\varphi|}, \quad I_c
= \int_0^{2\pi} \frac{\cos\varphi}{|\Delta-K\sin\varphi|}
\,d\varphi, \quad I_s = \int_0^{2\pi} \frac{\sin\varphi}{|\Delta-K\sin\varphi|} \,d\varphi
\end{equation}

Finally, after evaluating the integrals and substituting back in Eq.~\ref{eq:22} the closed form damping equation for phase slipping is given as

\begin{equation}
\label{eq:24}
c_a^{\mathrm{slip}}=\frac{\hat Q}{A\Omega}\frac{\Delta-\sgn(\Delta)\sqrt{\Delta^2-K^2}}{K}\sin\chi, \quad |\Delta|>K
\end{equation}

% At $|\Delta|=K$, Equations~\eqref{eq:lockedfinalgeneral} and \eqref{eq:slipfinalgeneral} have the same limiting value, so the two damping branches are continuous at the synchronisation boundary.

\subsection{Full Closed-Form Damping and Stiffness Equations and Small-Amplitude Approximations}

Combining the phase-locked and phase-slipping branches gives the complete first-harmonic aerodynamic damping expression:

\begin{equation}
\label{eq:25}
c_a=
\begin{cases}
-\dfrac{\widehat Q}{A\Omega}\left[\sqrt{1-\left(\dfrac{\Delta}{K}\right)^2}\cos\chi-\dfrac{\Delta}{K}\sin\chi\right], & |\Delta|\leq K, \\[9mm]
\dfrac{\widehat Q}{A\Omega}\dfrac{\Delta-\sgn(\Delta)\sqrt{\Delta^2-K^2}}{K}\sin\chi, & |\Delta|>K.
\end{cases}
\end{equation}

The equivalent aerodynamic stiffness follows from the component of the aerodynamic force in phase with the structural displacement. Since its derivation follows exactly the same procedure as that for the damping expression, it is given simply as

\begin{equation}
\label{eq:26}
k_a=
\begin{cases}
\dfrac{\widehat Q}{A}\left[\dfrac{\Delta}{K}\cos\chi+\sqrt{1-\left(\dfrac{\Delta}{K}\right)^2}\sin\chi\right], & |\Delta|\leq K, \\[9mm]
\dfrac{\widehat Q}{A}\dfrac{\Delta-\sgn(\Delta)\sqrt{\Delta^2-K^2}}{K}\cos\chi, & |\Delta|>K.
\end{cases}
\end{equation}

At $|\Delta|=K$, the two expressions in Eqs.~\ref{eq:25} and ~\ref{eq:26} have the same limiting value and therefore the phase-locked and phase-slipping branches are continuous at the lock-in boundary. To evaluate Eqs.~\ref{eq:25} and~\ref{eq:26} without forced-response calculations over the complete $(A,\Omega)$ parameter space, the following subsections exploit the first-order phase model and prescribed harmonic input to determine $K$ and $\Gamma_0$, while developing small-amplitude approximations for $\widehat Q$ and $\chi$.

\subsubsection{Small Amplitude Approximation of $\widehat Q$ and $\chi_Q$}

Under the small-amplitude approximation the first harmonic of the forced aerodynamic response is assumed to retain the amplitude and phase offset of the buffet response:

\begin{equation}
\label{eq:27}
Q_1(t;A,\Omega)=\widehat Q(A,\Omega)\cos\!\left[\theta+\chi_Q(A,\Omega)\right]\approx\widehat Q_B\cos\!\left(\theta+\chi_{Q,B}\right).
\end{equation}

\noindent Thus, the forced-response first-harmonic amplitude and phase offset are approximated by the unforced buffet values:

\begin{equation}
\label{eq:28}
\boxed{\widehat Q(A,\Omega)\approx\widehat Q_B, \qquad \chi_Q(A,\Omega)\approx\chi_{Q,B}.}
\end{equation}

\subsubsection{Harmonic Coupling Strength, Detuning, and Phase, $K$, $\Delta$, and $\chi$}

In the present case, the general forcing vector $\mathbf{f}$ reduces to a single scalar perturbation for the prescribed harmonic motion

\begin{equation}
\label{eq:29}
f(\phi;A)=A\widetilde{p}(\phi)=A\sin\phi
\end{equation}

\noindent where $\widetilde{p}(\phi)=\sin\phi$ is the unit-amplitude pitch waveform. The phase-coupling function therefore becomes

\begin{equation}
\label{eq:30}
\Gamma(\psi;A)=A\widetilde{\Gamma}(\psi),
\qquad
\widetilde{\Gamma}(\psi)
=
\frac{1}{2\pi}
\int_0^{2\pi}
Z(\phi+\psi)\widetilde{p}(\phi)\,d\phi
\end{equation}

\noindent where $\mathbf{Z}$ has also been replaced by the corresponding scalar phase-sensitivity function $Z$. Because $\widetilde{p}(\phi)$ contains a single harmonic, $\widetilde{\Gamma}(\psi)$ is exactly sinusoidal with zero mean, giving

\begin{equation}
\label{eq:31}
A\widetilde{\Gamma}(\psi)
=
A\widetilde{\Gamma}^{(1)}(\psi)
=
-Ak\sin\!\left(\psi+\widetilde{\beta}\right)
\end{equation}

\noindent where $k$ and $\widetilde{\beta}$ are the magnitude and phase of the first harmonic, respectively. Within the first-order phase model, comparison with Eq.~\ref{eq:8} then gives

\begin{equation}
\label{eq:33}
\boxed{
\Gamma_0=0,
\qquad
K(A)=Ak,
\qquad
\beta=\widetilde{\beta},
\qquad
\Delta=\omega_b-\Omega,
\qquad
\chi(A,\Omega)\approx\widetilde{\chi}
=
\chi_{Q,B}-\widetilde{\beta}
}
\end{equation}

The final closed form expressions for the phase reduced aerodynamic damping and stiffness under a small amplitude harmonic excitation are given by

\begin{equation}
\label{eq:34}
\boxed{
c_a=
\begin{cases}
-\dfrac{\widehat Q_B}{A\Omega}\left[\sqrt{1-\left(\dfrac{\Delta}{Ak}\right)^2}\cos \widetilde \chi-\dfrac{\Delta}{Ak}\sin \widetilde  \chi\right], & |\Delta|\leq Ak, \\[9mm]
\dfrac{\widehat Q_B}{A\Omega}\dfrac{\Delta-\sgn(\Delta)\sqrt{\Delta^2-(Ak)^2}}{Ak}\sin\widetilde\chi, & |\Delta|>Ak.
\end{cases}
}
\end{equation}

\begin{equation}
\label{eq:35}
\boxed{
k_a=
\begin{cases}
\dfrac{\widehat Q_B}{A}\left[\dfrac{\Delta}{Ak}\cos\widetilde  \chi+\sqrt{1-\left(\dfrac{\Delta}{Ak}\right)^2}\sin \widetilde  \chi \right], & |\Delta|\leq Ak, \\[9mm]
\dfrac{\widehat Q_B}{A}\dfrac{\Delta-\sgn(\Delta)\sqrt{\Delta^2-(Ak)^2}}{Ak}\cos \widetilde  \chi , & |\Delta|>Ak.
\end{cases}
}
\end{equation}

        \section{Computational Models}
    
     The numerical example for the present study has been performed for the ONERA OAT15A aerofoil, with experimental measurements available from the transonic wind tunnel of the Onera-Meudon Centre in France~\cite{jacquin09}. The experimental model is designed to study 2D buffet, with a chord length of $c = 0.23$m, a span of 0.78m and a thick trailing edge of 0.005$c$. Experiments were performed over a Mach number range of $0.70 \leq M_\infty \leq 0.75$ and a wind-off angle-of-attack (AOA) sweep of $2.4^\circ \leq \alpha_0 \leq 3.91^\circ$ to determine the transonic buffet envelope onset at a Reynolds number of $Re_\infty = 3.0\times10^6$ (based on the chord length). 
     
\subsection{Computational Fluid Dynamics Model}
    The general purpose finite volume code ANSYS Fluent 2025 R2~\cite{ansys} is used. The 2D URANS equations are solved using the pressure-based implicit solver, with a second-order upwind scheme for the convective terms, central differencing for the diffusive terms, and Rhie--Chow distance-based interpolation for the face mass fluxes. A dual time-stepping scheme is employed with second-order implicit temporal discretisation. Temporal convergence studies are conducted with non-dimensional time-steps of $\Delta \tau = t(u_\infty/c) = 0.01, \, 0.005, \, 0.0025$, where $u_\infty$ is the freestream velocity. The Generalized $k-\omega$ turbulence model~\cite{menter21} is used with the Spalart-Shur curvature correction~\cite{spalart97}. The convergence criteria are set to $1\times10^{-5}$ for the scaled residuals at each time-step. Committee-supplied structured C-grid topologies are used, which are available for download~\footnote{https://aiaa-dpw.larc.nasa.gov/grids.html}. The grids extend 150 chord lengths in all directions. Three grids of varying refinement are used in a combined spatial and temporal convergence study, with the mesh statistics, convergence results, and GEKO-model calibration and validation provided in Appendix~\ref{appA}.

\begin{figure}[h!]
  \centerline{\includegraphics[width=0.4\textwidth]{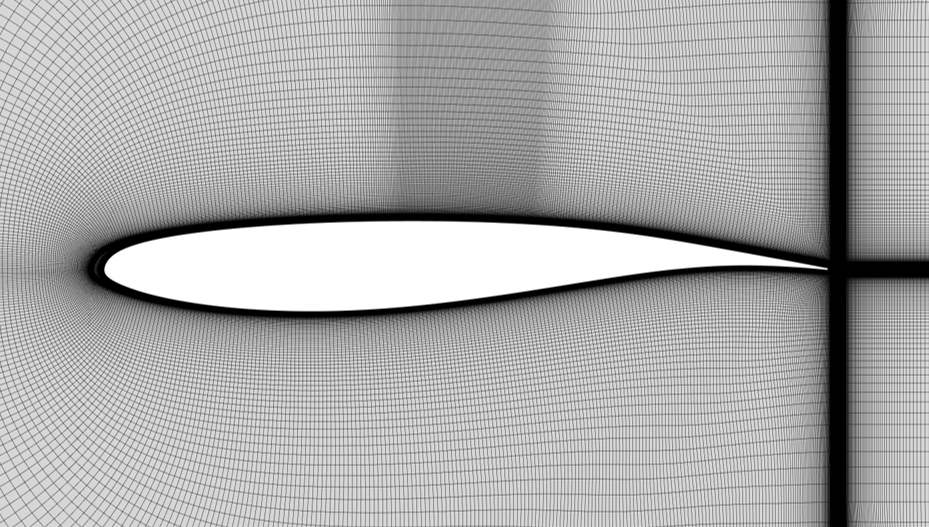}}
  \caption{Computational grid around the aerofoil (L3).}
  \label{fig:mesh}
\end{figure}

\subsection{Time-Domain Aerodynamic Reduced Order Model}
\label{sec:time_domain_rom}

To verify the solutions of the closed-form phase reduction model, alongside coupled time-domain CFD solutions, a time-domain aerodynamic ROM is used. The reason for this model is that the generation of aerodynamic work maps in the time-domain requires prescribed harmonic motion CFD runs across a dense amplitude–frequency space, with phase slipping cases requiring millions of time steps to converge. Therefore, although expensive to train, the time-domain nonlinear ROM retains the nonlinear, amplitude-dependent aerodynamic response and can provide a high-fidelity reference in the low-to-moderate amplitude regime. 

\begin{figure}[h!]
  \centerline{\includegraphics[width=\textwidth]{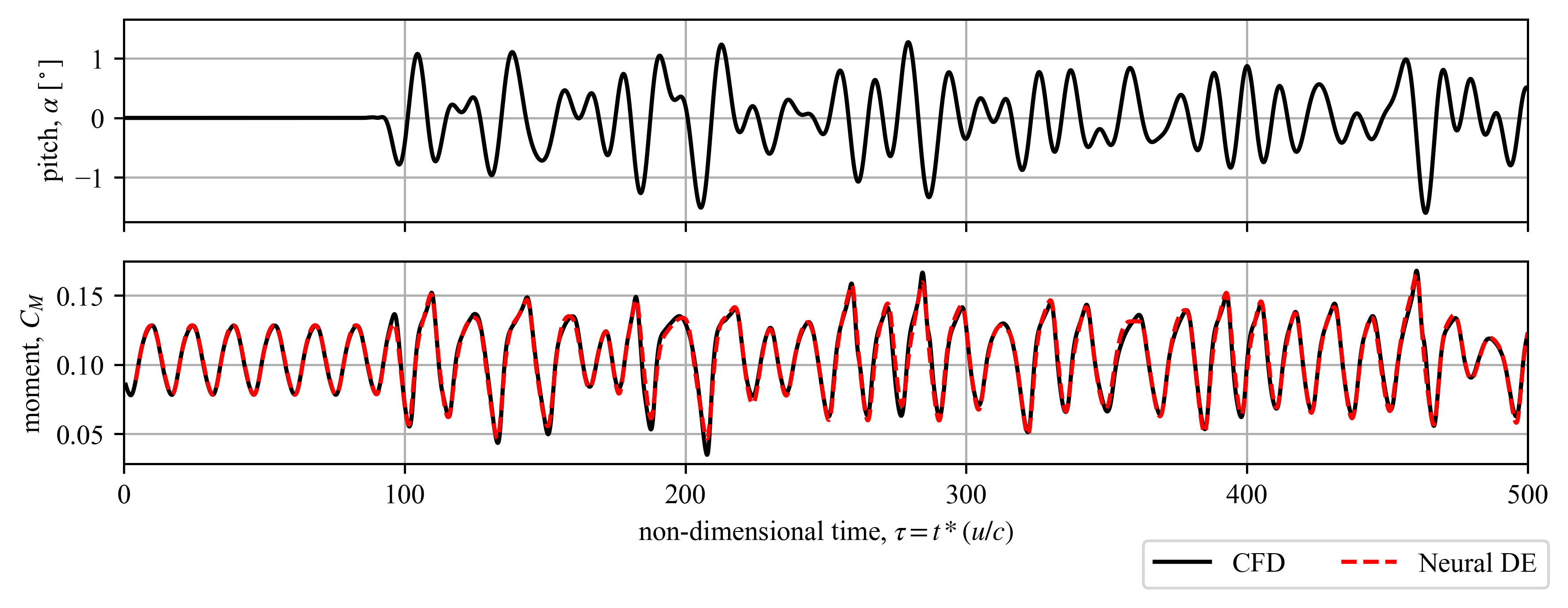}}
  \caption{Cross-validation of the pitch-moment time-domain aerodynamic ROM.}
  \label{fig:rom_cross_validation}
\end{figure}

The ROM is described briefly here and follows the physics-guided Neural differential equation (DE) formulation~\cite{chen18,rackauckas20,kidger20} developed and extensively verified for transonic buffet in the authors' prior work, including for forced response and aeroelastic predictions~\cite{candon26b}. It is shown to be accurate and highly consistent over the low-to-moderate amplitude range considered here, as confirmed by the pitch and heave cross-validation results summarised in Table~\ref{tab:rom_cross_validation} and illustrated in Fig.~\ref{fig:rom_cross_validation}. Rather than writing the general formulation, it will be given in the form that is used in this paper. Let $\tilde q_n$ denote either the generalised structural displacement or velocity, and $\tilde q_{n-k}$ denote its current and lagged values, then model for the generalised aerodynamic force $Q_n$ is given as

\begin{equation}
\label{eq:time_domain_rom}
\begin{aligned}
\dot{Q}_n &= V_n \\
\dot{V}_n &=
\underbrace{
\epsilon\left(1-\alpha V_n^{2}\right)V_n
-\omega_B^2Q_n
}_{\text{Rayleigh oscillator}}
+
A\ddot{q}_n
+
\underbrace{
\sum_{k=0}^{N_L}H^{\tilde q}_k\tilde{q}_{n-k}
}_{\text{Volterra forcing}}
+
\underbrace{
\mathcal{F}_{\mathrm{NN}}\left(Q_n,V_n,\mathbf{z}_n\right)
}_{\text{neural correction}}
\end{aligned}
\end{equation}

% ALet $\boldsymbol{\tilde q}_n=[\tilde{q}_{1,n},\ldots,\tilde{q}_{M,n}]^{\mathsf T}$ represent the generalised structural displacement, $\boldsymbol{\tilde q}_n = \boldsymbol{q}_n$, or velocity, $\boldsymbol{\tilde q}_n = \boldsymbol{\dot{q}}_n$, and let $Q_{i,n}$ represent the generalised aerodynamic force output, then the model is

% \begin{equation}
% \label{eq:time_domain_rom}
% \dot{Q}_{i,n} = V_{i,n} 
% \dot{V}_{i,n}=
% \epsilon_i\left(1-\alpha_iV_{i,n}^{2}\right)V_{i,n}
% -\omega_F^2Q_{i,n}
% +\ddot{q}_{j,n} 
% \quad+
% \sum_{k=0}^{N_L}
% H_{i,\tilde q}^{j}[k]\tilde{q}_{j,n-k}
% +
% \mathcal{F}_{\mathrm{NN},i}
% \left(Q_{i,n},V_{i,n},\mathbf{z}_n\right)
% \end{equation}

\noindent where a Rayleigh oscillator represents the autonomous buffet dynamics~\cite{hartlen70,dowell81}, a first-order Volterra kernel $\mathbf H^{\tilde q}$ represents the finite-memory response to structural motion~\cite{silva97,balajewicz12}, and $\mathcal{F}_{\mathrm{NN}}()$ is a small neural-network correction which takes the aerodynamic states ($Q_n$, $V_n$) and the current and lagged generalised structural inputs $\mathbf z_n$. 

The model parameters are identified from prescribed-motion CFD data through backpropagation over the complete discrete-time rollout. Separate ROMs are used for prescribed pitch and heave motion, while the ROM is not used in the two-degree-of-freedom analysis. For the pitch ROM, the input is the pitch displacement, $\alpha$, whereas for the heave ROM it is the heave rate, $\dot{h}$. In both cases, the training and cross-validation data are generated using band-limited noise over the frequency ranges reported in Table~\ref{tab:rom_cross_validation}. The datasets are downsampled by a factor of 10 relative to the CFD time step before training. The number of training samples $N_{\mathrm{train}}$, validation samples $N_{\mathrm{val}}$, and lags $N_L$, are defined based on the raw CFD data.

\begin{table}[!h]
	\centering
	\caption{Training parameters and cross-validation errors for the time-domain aerodynamic ROMs.}
    \label{tab:rom_cross_validation}
	\begin{tabular}{lcccccccc}
		\hline
		Input & Aero force & $N_{\mathrm{train}}$ & $N_{\mathrm{val}}$ & $\hat{f}$ range & Amplitude (max) & $N_L$ & HU & NRMSD [\%] \\
		\hline
		Heave, $h/b$ & $L$ & 100k & 100k & 0.5--1.5 & $\pm 0.1$ & 500 & 16 & 1.85 \\
		Pitch, $\alpha$ ($^\circ$) & $M_{c/4}$ & 100k & 100k & 0.5--1.5 & $\pm 1.5$ & 500 & 16 & 2.01 \\
		\hline\hline
	\end{tabular}
\end{table}

\clearpage

\subsection{Aeroelastic Equations of Motion}

A two-dimensional typical-section structural model is used with heave, $h$, and pitch, $\alpha$, degrees of freedom. The structural equations are written as

\begin{equation}
\mathbf{M}\ddot{\mathbf{x}}+\mathbf{C}\dot{\mathbf{x}}+\mathbf{K}\mathbf{x}=\mathbf{F}_a
\label{eq:ae1}
\end{equation}

\noindent where $\mathbf{x}=[h,\alpha]^{\mathsf T}$ and $\mathbf{F}_a=[L,M_{c/4}]^{\mathsf T}$. For the present typical section used here, these are given as

\begin{equation}
\begin{bmatrix}
m & mx_\alpha b\\
mx_\alpha b & I_\alpha
\end{bmatrix}
\begin{bmatrix}
\ddot{h}\\
\ddot{\alpha}
\end{bmatrix}
+
\begin{bmatrix}
c_{hh} & c_{h\alpha}\\
c_{h\alpha} & c_{\alpha\alpha}
\end{bmatrix}
\begin{bmatrix}
\dot{h}\\
\dot{\alpha}
\end{bmatrix}
+
\begin{bmatrix}
k_h & 0\\
0 & k_\alpha
\end{bmatrix}
\begin{bmatrix}
h\\
\alpha
\end{bmatrix}
=
\begin{bmatrix}
L\\
M_{c/4}
\end{bmatrix}
\label{eq:ae2}
\end{equation}

\noindent where $m$ is the sectional mass, $I_\alpha$ is the sectional moment of inertia about the elastic axis, and $x_\alpha b$ is the offset between the centre of mass and elastic axis. The baseline structural-to-fluid mass ratio is $\mu=m/(\pi\rho b^2)=870$, where $b=0.115$~m and $\rho=0.923$~kg/m$^3$. The elastic axis is located at $x/c=0.25$. The structural natural frequencies are denoted by $\omega_1$ and $\omega_2$, and structural natural frequency ratios by $\hat \omega_1 = \omega_1 / \omega_B$ and $\hat \omega_2 = \omega_2 / \omega_B$. 

The physical and modal coordinates are related by $\mathbf{x}=\boldsymbol{\Phi}\mathbf{q}$, where $\boldsymbol{\Phi}$ contains the structural mode shapes and $\mathbf{q}$ contains the modal coordinates. Substitution into Eq.~\ref{eq:ae1} and pre-multiplication by $\boldsymbol{\Phi}^{\mathsf T}$ gives

\begin{equation}
\ddot{\mathbf{q}}+\mathbf{C}_q\dot{\mathbf{q}}+\mathbf{K}_q\mathbf{q}=\mathbf{Q}
\label{eq:ae3}
\end{equation}

\noindent where mass-normalised modes are used such that $\boldsymbol{\Phi}^{\mathsf T}\mathbf{M}\boldsymbol{\Phi}=\mathbf{I}$, $\mathbf{C}_q=\boldsymbol{\Phi}^{\mathsf T}\mathbf{C}\boldsymbol{\Phi}$, $\mathbf{K}_q=\boldsymbol{\Phi}^{\mathsf T}\mathbf{K}\boldsymbol{\Phi}=\operatorname{diag}(\omega_1^2,\omega_2^2)$, and $\mathbf{Q}=\boldsymbol{\Phi}^{\mathsf T}\mathbf{F}_a$. In the frequency domain, the generalised aerodynamic forces are represented by equivalent modal aerodynamic damping and stiffness matrices, giving

\begin{equation}
\ddot{\mathbf{q}}
+
\left[
\mathbf{C}_q+s\mathbf{C}_a(\Omega)
\right]\dot{\mathbf{q}}
+
\left[
\mathbf{K}_q+s\mathbf{K}_a(\Omega)
\right]\mathbf{q}
=
\mathbf{0}
\label{eq:ae4}
\end{equation}

\noindent where $\mathbf{C}_a(\Omega)$ and $\mathbf{K}_a(\Omega)$ are the frequency-dependent modal aerodynamic damping and stiffness matrices, respectively, which are obtained from the closed-form expressions given in Eqs.~\ref{eq:34} and~\ref{eq:35} (derived in the previous Section~\ref{sec:derivation}), and $s$ is a linear scaling of the aerodynamic forces. Assuming $\mathbf{q}(t)=\widehat{\mathbf{q}}e^{pt}$ gives the quadratic eigenvalue problem

\begin{equation}
\left\{
p^2\mathbf{I}
+
p\left[
\mathbf{C}_q+s\mathbf{C}_a(\Omega)
\right]
+
\left[
\mathbf{K}_q+s\mathbf{K}_a(\Omega)
\right]
\right\}
\widehat{\mathbf{q}}
=
\mathbf{0}
\label{eq:ae5}
\end{equation}

\noindent where $p=\sigma+\mathrm{i}\Omega$. Equation~\ref{eq:ae5} is solved using a $p$-$k$-type fixed-point iteration. At each iteration, the aerodynamic matrices are evaluated at the current frequency, the quadratic eigenvalue problem is solved, and the frequency is updated from the imaginary part of the selected eigenvalue until convergence. The damping ratio is calculated as

\begin{equation}
\zeta=-\frac{\Re(p)}{|p|}
\label{eq:ae6}
\end{equation}

\noindent such that flutter occurs when $\zeta=0$. For the finite-amplitude analysis, the same procedure is repeated using $\mathbf{C}_a(\Omega,A)$ and $\mathbf{K}_a(\Omega,A)$ to estimate the aeroelastic bifurcation diagram.

\section{Results}

The results considering the baseline OAT15A operating conditions of $ M_\infty = 0.73$ and wind-off AOA of $ \alpha_0 = \leq 3.5^\circ$. Unless otherwise specified, the baseline Reynolds number is $Re_\infty = 3.0\times10^6$ (based on the chord length). The results are separated into three parts:

\begin{enumerate}
    \item The PSFs and PCFs are presented and discussed. 
    \item The Arnold tongues and predictions of the closed-form damping equations are analysed.
    \item The two-degree-of-freedom frequency domain aeroelastic solutions are analysed.
\end{enumerate}

\subsection{Phase-Sensitivity and Phase-Coupling Functions}

\begin{figure}[h!]
    \centering
    \includegraphics[width=1.0\textwidth]{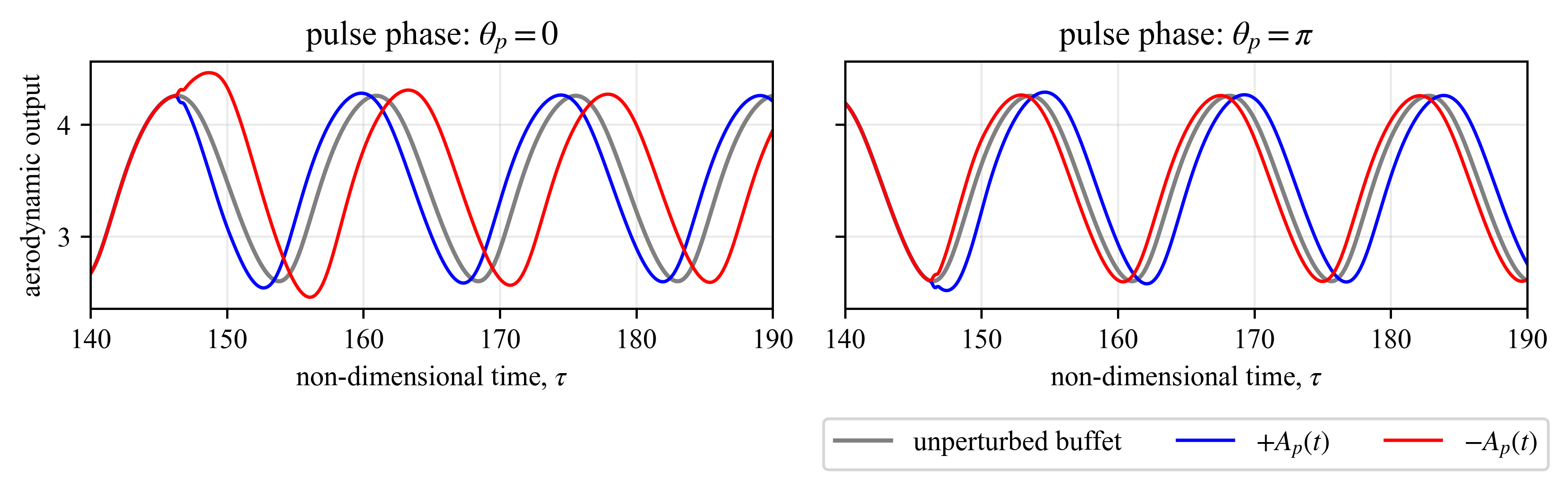}
    \caption{Representative aerodynamic responses following exaggerated positive and negative pitch pulses. }
    \label{fig:pulse_phase_response}
\end{figure}

\begin{figure}[!h]
    \centering
    \subfigure[Heave--lift]{%
    \label{fig:phase_functions_heave}
    \includegraphics[width=0.95\textwidth]{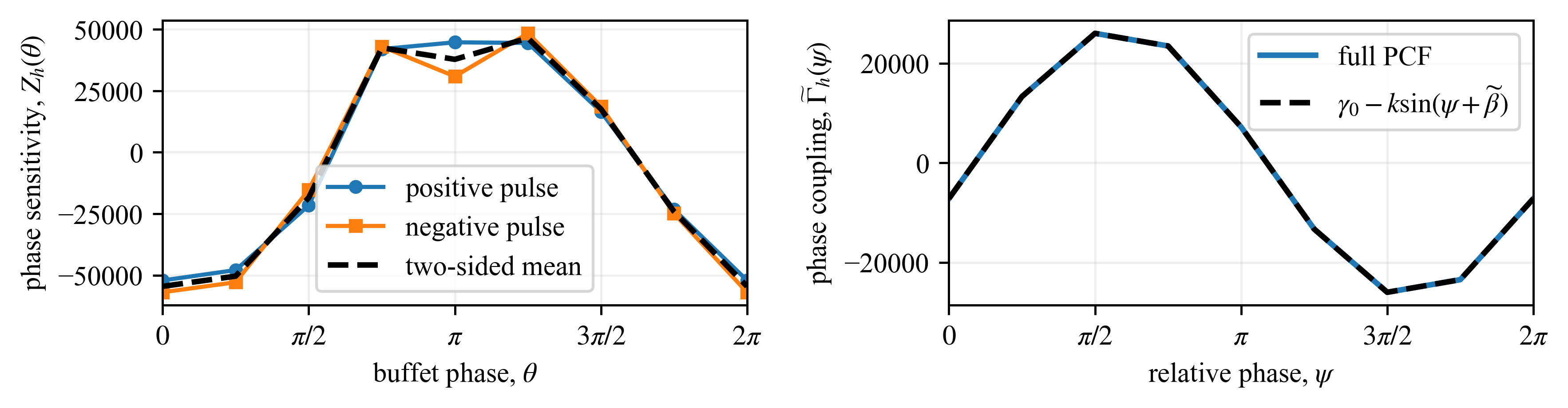}}

    \subfigure[Pitch--moment]{%
    \label{fig:phase_functions_pitch}
    \includegraphics[width=0.95\textwidth]{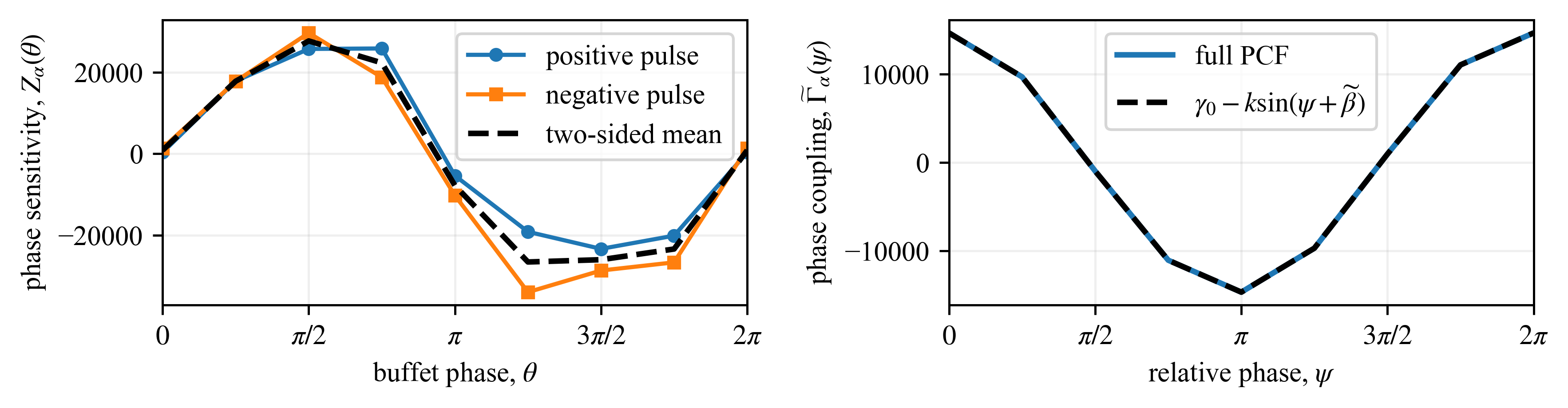}}

    \caption{Phase-sensitivity and phase-coupling functions with the
    reference phase based on the maximum positive moment.}
    \label{fig:phase_functions}
\end{figure}

The PSF is obtained by applying short pulses through the structural coordinate at prescribed phases, $\theta_p$, of the buffet limit cycle. Following each pulse, the perturbed aerodynamic response is allowed to settle towards the stable buffet orbit. Provided that the perturbation is weak, the effect is a shift in the phase of the buffet oscillation rather than a change in its amplitude or waveform. Figure~\ref{fig:pulse_phase_response} shows exaggerated examples of positive and negative pitch pulses applied at two different phases of the buffet cycle. Depending on $\theta_p$ and the sign of the perturbation, the buffet oscillation may advance or retreat relative to the unperturbed limit cycle. The phase is far less sensitive than the exaggerated examples in Fig.~\ref{fig:pulse_phase_response} makes it seem. As is shown next in Fig.~\ref{fig:phase_functions}, under the the small amplitude pulses required to stay within the linear regime, the direction becomes substantially less important (halving the number of simulations required to perform the phase reduction).

After the transient response has decayed, the asymptotic phase shift, $\Delta\theta_\infty(\theta_p)$, produced by a pulse of area $I$ is measured. The corresponding value of the PSF is then calculated using Eq.~\ref{eq:res1} if pulses are applied in one direction, or using Eq.~\ref{eq:res2} a two-sided central-difference estimate. Repeating this procedure over the complete buffet cycle gives the full PSF, $Z(\theta)$, and then the PCF, $\widetilde{\Gamma}(\psi)$, can be obtained using Eq.~\ref{eq:30}, followed by the conversion to Adler form, $\widetilde{\Gamma}^{(1)}(\psi)$, as per Eqs.~\ref{eq:31} and~\ref{eq:33}.

\begin{equation}
\label{eq:res1}
Z_\alpha(\theta_p)=\frac{\Delta\theta_\infty(\theta_p)}{I}
\end{equation}

\begin{equation}
\label{eq:res2}
Z_\alpha(\theta_p)=\frac{\Delta\theta_\infty(\theta_p)^+-\Delta\theta_\infty(\theta_p)^-}{2I}
\end{equation}

Using the maximum positive pitching moment as the common reference phase, $\theta$, Fig.~\ref{fig:phase_functions} shows the heave and pitch PSFs, $Z_h(\theta)$ and $Z_\alpha(\theta)$, with the corresponding PCFs, $\widetilde{\Gamma}_h(\psi)$ and $\widetilde{\Gamma}_\alpha(\psi)$. The PSFs are obtained using pulses with a duration of 0.2 of the buffet period and peak amplitudes of $A_{p,(h/b)} = 8.7\times10^{-4}$ ($A_{p,h} = 1\times10^{-4}$m) and $A_{p,\alpha} = 0.005^\circ$ for the heave and pitch perturbations, respectively. The linear range of pitch pulses was found to be $A_{p,\alpha} \lesssim 0.01^\circ$, while $A_{p,\alpha} \gtrsim  0.002^\circ$ is required to provide sufficient excitation above the numerical noise of the URANS solution. The buffet limit cycle is discretised by $N_{\theta_p} = 8$ pulses evenly spaced between $0 \leq \theta_p \leq 2\pi$. After the pulse, the simulation is run for an additional $N_c = 8$ cycles and the phase is measured from the last six cycles. Sensitivity analysis and justification of these parameters is provided in Appendix~\ref{appB}, along with evidence to suggest that they can be substantially reduced for the purposes of aeroelastic modelling. 

The lift and moment first harmonics are nearly in phase (lift leads moment by approximately $2^\circ$) which means that expressing the heave--lift channel using the moment-based global phase is effectively equivalent to using its own lift-based phase. This is relevant to the aeroelastic interpretation of the results because aeroelastic modelling requires consistent work-conjugate generalised coordinate--force pairs.

The results show that the lift and moment are most sensitive to heave perturbations when applied near the extrema of their cycles, while for pitch perturbations the greatest sensitivity occurs near where the aerodynamic loads experience a maximum gradient. A positive sign implies that the perturbation advances the buffet phase, while a negative sign implies that it delays the phase. Of particular practical relevance for flow control, this means that a heave perturbation has maximal authority when applied near the extrema of the shock cycle, corresponding approximately to its maximum upstream or downstream excursion, whereas a pitch perturbation has maximal authority near the phases of maximum shock-motion gradient.

\subsection{Arnold Tongue and Closed-Form Aerodynamic Damping}
\begin{figure}[h!]
	\centering
		\includegraphics[width=1.0\textwidth]{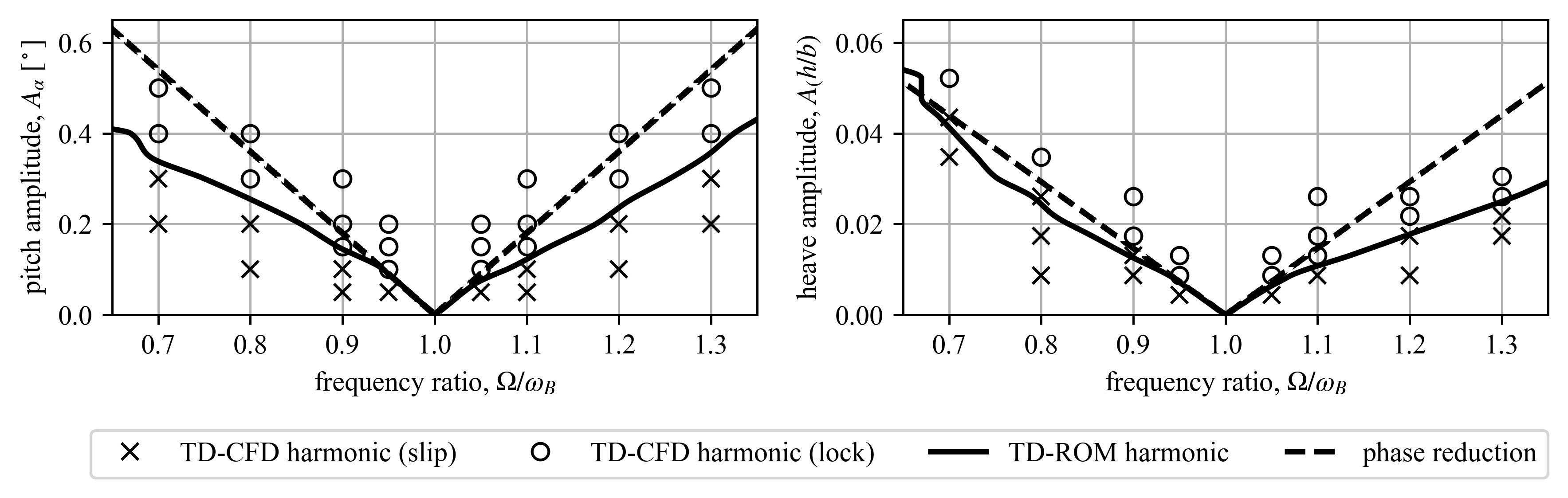}
	\caption{Arnold tongues predicted using various methods.}
	\label{fig:arnold}
\end{figure}

The Arnold tongue describes the lock-in boundary of the buffet cycle as a function of the harmonic frequency and amplitude through which it is excited. The Arnold tongues as predicted using the time-marching URANS code (TD-CFD harmonic), the time-marching ROM (TD-ROM harmonic), and by phase reduction are presented in Fig.~\ref{fig:arnold}. As is evident in the time marching CFD and ROM results, there is a significant nonlinear amplitude dependency of the tongue as the excitation frequency departs from $\Omega/\omega_B = 1$, which is not unexpected in the case of shock buffet. The phase reduction model used here is based osn small amplitude approximations and does not capture this nonlinear amplitude dependency, demonstrating accurate predictions for heave up to $A_{(h/b)}\approx0.01-0.02$ ($\sim1\%-2\%$ of the semi-chord), and for pitch up to $A_\alpha\approx0.1^\circ-0.2^\circ$. While it may seem that the limits of small amplitude phase reduction are being pushed here, it will be shown that having an accurate model in the small amplitude limit and an approximate model at larger amplitudes is both useful and insightful.

Obtaining enough time-domain data to estimate the aerodynamic damping in the slipping condition requires the simulation of millions of physical time steps for each combinations of $A$ and $\Omega$. Therefore, having verified sufficient accuracy of the time-domain ROM (Table~\ref{tab:rom_cross_validation} and Fig.~\ref{fig:arnold}), it is used as a surrogate for the time-marching CFD in the remainder of the section to verify the closed form aerodynamic models proposed in Eq.~\ref{eq:34} and~\ref{eq:35}. 

% Aerodynamic work maps for the pitch--moment channel, obtained either directly from the long-time average of the time-marching ROM data (left) or from Eq.~\ref{eq:34} through $W_{\mathrm{aero}}=-\pi A^2\Omega c_a$ (right), are presented in Fig.~\ref{fig:pitch_work}.

\begin{figure}[h!]
	\centering
		\includegraphics[width=1.0\textwidth]{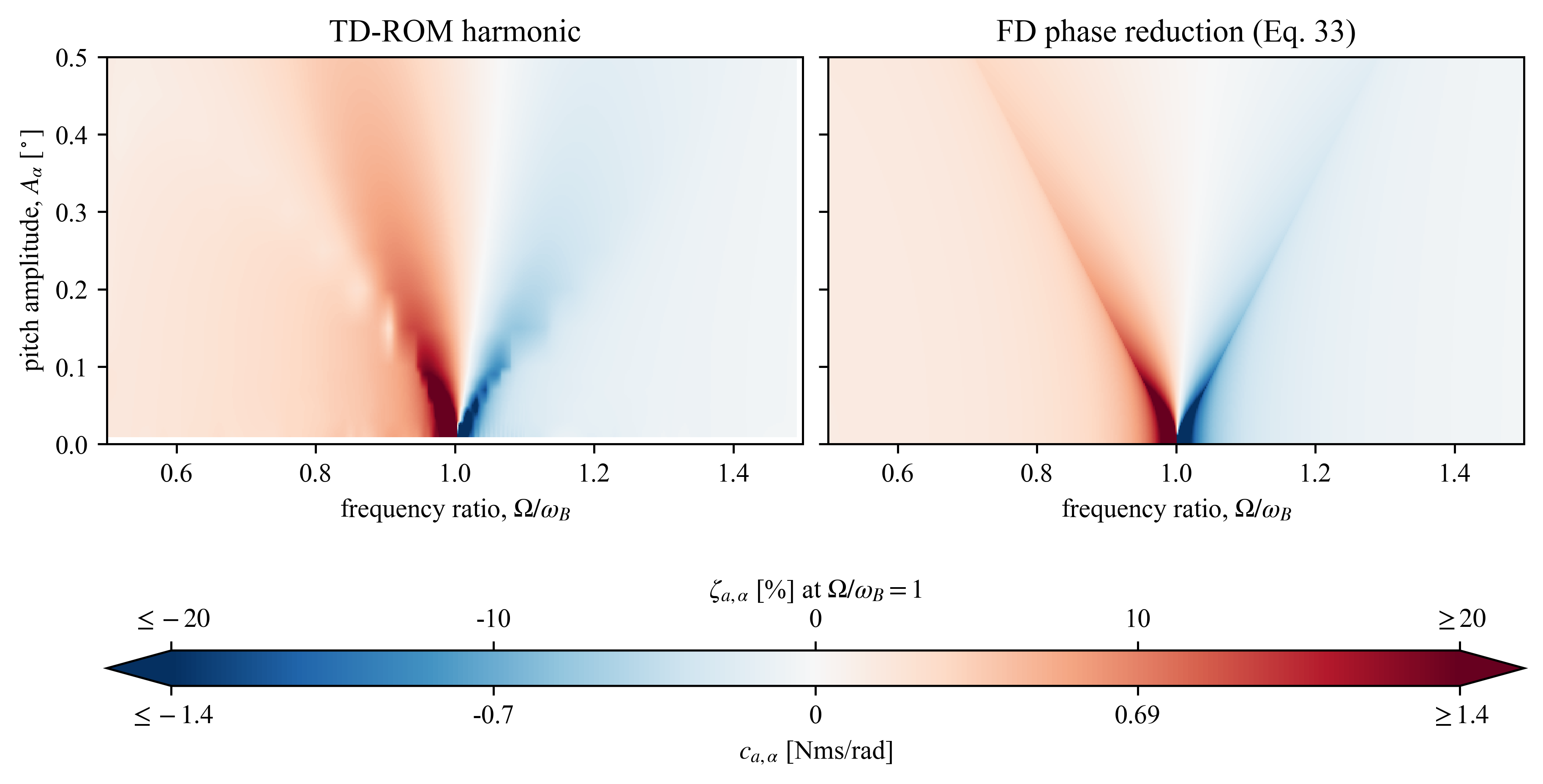}
	\caption{Aerodynamic damping maps for the pitch-moment channel.}
	\label{fig:aerodamping_A}
\end{figure}

Aerodynamic damping maps for the pitch--moment channel obtained from the long-time average of the time-marching ROM data, and directly from the derived frequency domain closed form damping Eq.~\ref{eq:34} (FD-phase reduction), are presented in Fig.~\ref{fig:aerodamping_A}. Negative aerodynamic damping (positive work) indicates that the aerodynamic loading injects energy into the structure and is destabilising. Strong qualitative agreement between the time marching ROM and the closed form model is observed, demonstrating that the aerodynamic loading is predominantly destabilising for frequency ratios $\Omega/\omega_B>1$, entirely consistent with numerous prior aeroelastic studies~\cite{raveh14,giannelis16,gao17,candon26b}. The aerodynamic damping ratio $\zeta_{a,\alpha}=c_{a,\alpha}/(2I_\alpha\Omega_R)$ is evaluated using $\Omega_R=\omega_B \, \forall \, \Omega$, allowing direct conversion between the two colour scales. Very large damping ratio values and an abrupt change of sign can be observed in the small amplitude limit for $\Omega/\omega_B \approx 1$, which is attributed to the $1/A$ aerodynamic damping singularity predicted by Eqs.~\ref{eq:17} and~\ref{eq:18}.

% \begin{figure}[h]
% 	\centering
% 		\includegraphics[width=1.0\textwidth]{Figures/JFM_pitch_work.PNG}
% 	\caption{Aerodynamic work maps for the pitch-moment channel.}
% 	\label{fig:pitch_work}
% \end{figure}

\begin{figure}[h!]
    \centering
    \subfigure[Heave--lift]{%
    \label{fig:aerodamping_B_heave}
    \includegraphics[width=0.95\textwidth]{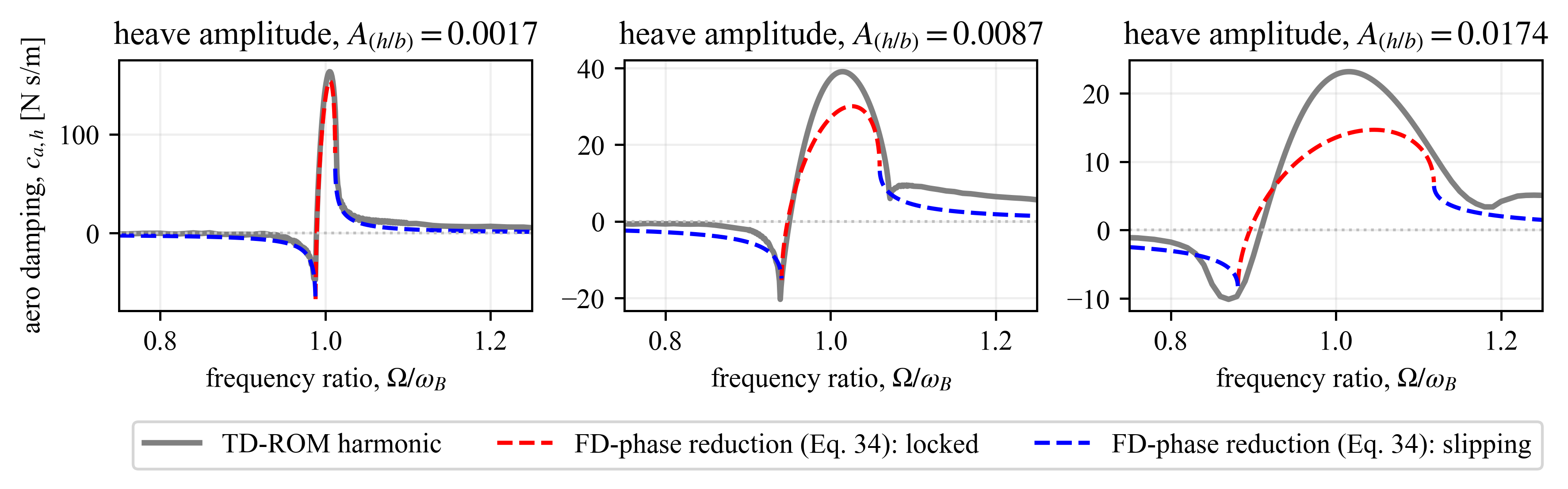}}

    \subfigure[Pitch--moment]{%
    \label{fig:aerodamping_B_pitch}
    \includegraphics[width=0.95\textwidth]{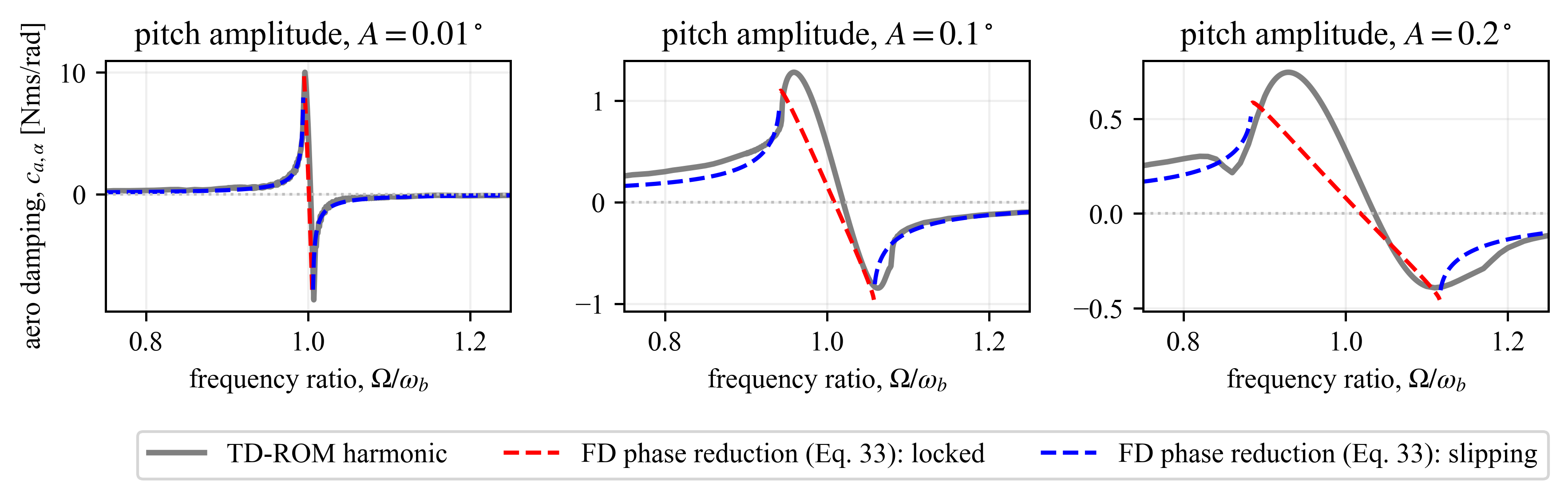}}

    \caption{Aerodynamic damping predictions for various harmonic amplitudes:
    (\textit{a}) heave--lift and (\textit{b}) pitch--moment.}
    \label{fig:aerodamping_B}
\end{figure}

Figure~\ref{fig:aerodamping_B} presents the damping predictions for the heave--lift and pitch--moment channels at various fixed amplitudes. For the heave--lift channel, negative aerodynamic damping occurs for frequency ratios $\Omega/\omega_B<1$ which is again consistent with prior aeroelastic studies~\cite{candon26b}. The $1/A$ aerodynamic damping singularity is captured very accurately by the closed-form model, as is the aerodynamic damping more generally in the small-amplitude limit. In both cases, as $A$ increases, the locked branch widens and the magnitude of the damping reduces substantially. The strongest negative damping generally occurs near the intersection of the locked and slipping branches. Although the closed-form model progressively loses pointwise accuracy as the amplitude increases, the location of the dominant negative damping region and the phase-slipping--phase-locked transition are well captured. As shown in the following subsection, this is an important aeroelastic feature that is tied to the amplitude of the subcritical fold point.
\clearpage

\begin{figure}[h!]
	\centering
		\includegraphics[width=0.98\textwidth]{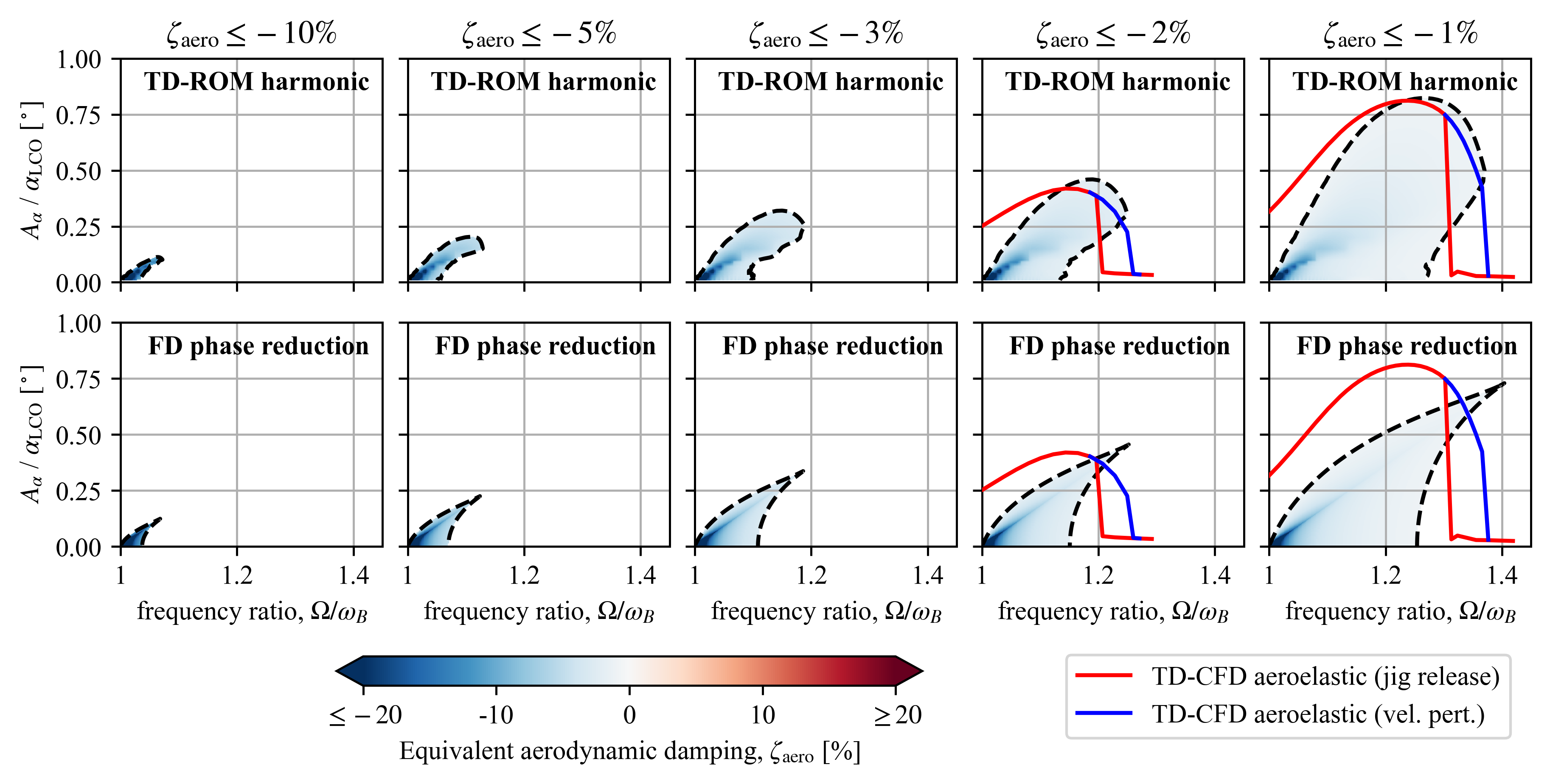}
	\caption{Aerodynamic damping ratio maps demonstrating structural damping exceedance.}
	\label{fig:aerodamping_C}
\end{figure}

From an aeroelastic perspective, the question is whether the negative aerodynamic damping is sufficient to overcome the structural damping. Given typical structural damping estimates of $\zeta_s \approx 2\%$, the magnitude of the negative damping predicted here is substantial and indicates that an aeroelastic LCO is unavoidable near the buffet frequency. From both a fundamental and practical perspective, the questions remain: how close do the fluid and structural natural frequencies need to be, how severe will the LCO be, and what is its character. Figure~\ref{fig:aerodamping_C} identifies the regions in which the negative aerodynamic damping ratio exceeds selected threshold values (corresponding to an equivalent structural damping). The aerodynamic damping ratio $\zeta_{a,\alpha}=c_{a,\alpha}/(2I_\alpha\Omega_R)$ is evaluated using $\Omega_R=\Omega$. Generally speaking, the dominant features of these curves include both linear and subcritical instability regions for $\Omega/\omega_B>1$. Although the infinitesimal amplitude linear stability boundary is difficult to recover exactly from a time marching aeroelastic simulation, the predicted aeroelastic LCO amplitudes, $\alpha_{\textrm{LCO}}$, show that both the stability range and LCO amplitude are strongly correlated with the harmonic zero-net-damping contours.\footnote{\textbf{jig release}: model is released from its wind-off shape and perturbed only by the natural aerodynamic loading. \textbf{vel. pert.}: large critical structural damping is applied for several time steps, followed by a release to nominal structural damping and a large velocity perturbation.}

\begin{figure}[h!]
	\centering
		\includegraphics[width=0.625\textwidth]{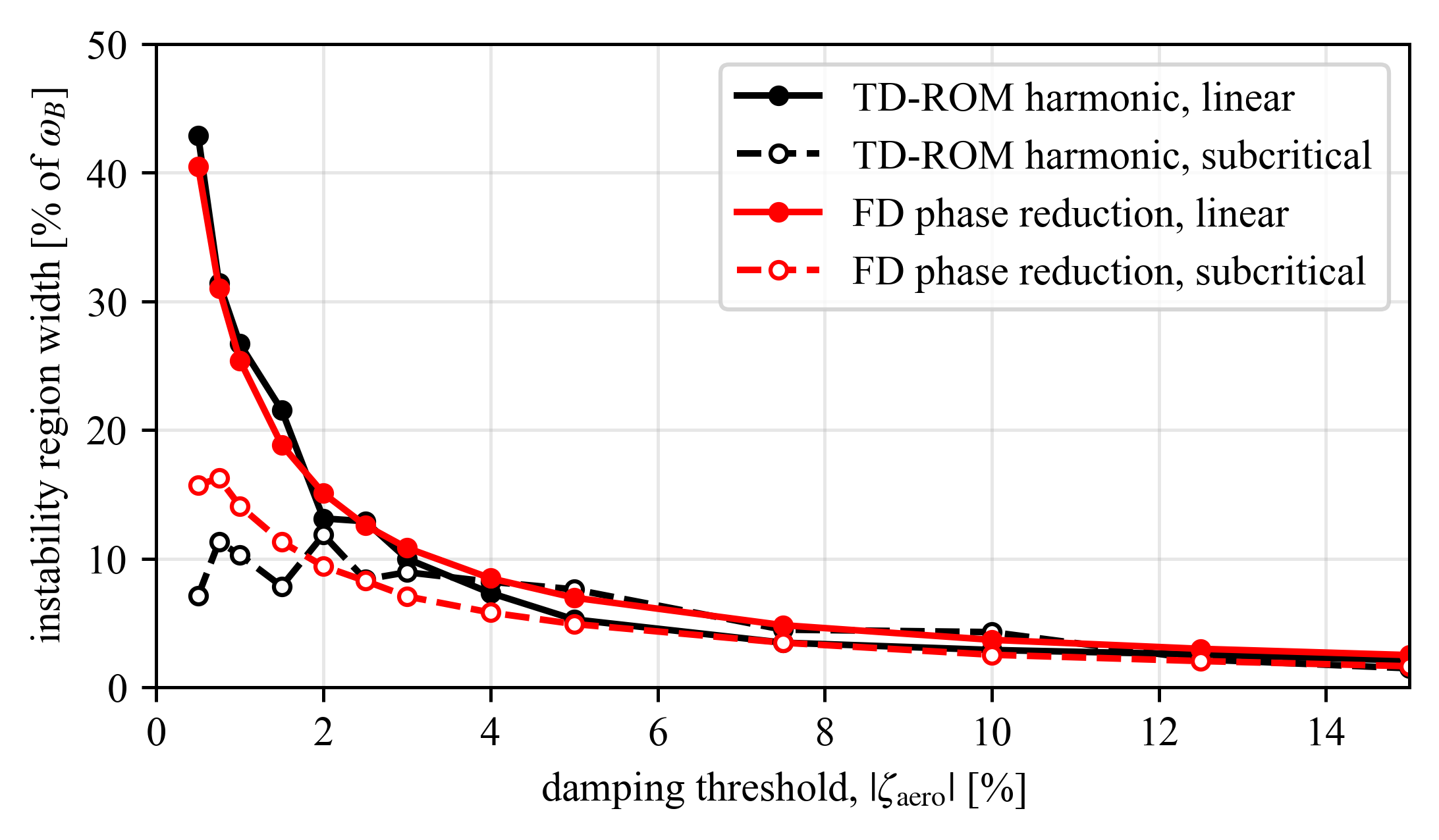}
	\caption{Linear and subcritical instability range versus damping threshold.}
	\label{fig:aerodamping_D}
\end{figure}

The widths of the linear and subcritical instability regions for selected damping thresholds are presented in Fig.~\ref{fig:aerodamping_D}, given as a percentage of the buffet frequency, $100(\Omega/\omega_B - 1)$. The width of the linear instability region increases rapidly as the damping threshold is reduced and approaches zero asymptotically as the threshold increases due to the $1/A$ singularity in Eq.~\ref{eq:34}, i.e., the region becomes arbitrarily small but never disappears. For the finite damping threshold values shown here the phase reduced model is able to recover the width of the linear instability region to within $2\%-3\%$ of the nonlinear time-domain ROM. However, what is not shown is that the linear instability range predicted by the phase reduced model grows unbounded as the damping threshold approaches zero (because the phase reduced model predicts that the negative aerodynamic damping approaches zero only asymptotically with increasing frequency ratio in the infinitesimal amplitude limit). In reality, the zero-damping instability range is finite, extending to approximately 90\% above the buffet frequency (for this case). This is predicted by the nonlinear time-domain ROM, and could also be predicted by extending the proposed method to a phase--amplitude reduction.  

The width of the subcritical region decreases alongside the linear range as the damping threshold increases beyond approximately $|\zeta_{\textrm{aero}}| \approx 2\%$. In this regime, the phase-reduced model predicts the subcritical range with high accuracy, exhibiting errors of only $2$--$3\%$, because the instability exists only at small harmonic amplitudes. However, for smaller damping thresholds, the subcritical range becomes non-monotonic, meaning that the relative linear-to-subcritical width grows rapidly. Here too, the subcritical range exists through much higher amplitudes, therefore the error in the phase reduced model increases to a maximum of approximately $10\%$  

\subsection{Frequency Domain Aeroelastic Predictions of a Two-Degree-of-Freedom Model}

Frequency-domain aeroelastic predictions are now obtained using the closed-form phase reduced aerodynamic model, with the aerodynamic matrices constructed from Eqs.~\ref{eq:34} and~\ref{eq:35}. To the authors' knowledge, this represents the first frequency-domain aeroelastic predictions of this nature for transonic buffet and the first investigation of the associated criticality as a function of an aerodynamic scaling parameter. For the two-degree-of-freedom system considered, $\omega_1$ and $\omega_2$ are the mode 1 and 2 structural natural frequencies respectively, and  $\hat \omega_1$ and $\hat \omega_2$ are their buffet-normalised values (natural frequency ratios). The time-domain ROM is not used as a reference; instead, full time-marching CFD-based aeroelastic computations are used to ensure a rigorous assessment of the phase reduction model.

For the phase reduction model, under the assumption of insensitivity to Reynolds number, the generalised forces scale linearly with dynamic pressure. The parameter $s$ linearly scales the aerodynamic forces relative to the reference condition at which the phase reduction was performed. For example, $s=2$ corresponds to a doubling of the dynamic pressure and the generalised forces from the reference condition. This is a standard approach to scale the forces in aeroelasticity. However, in reality for $Re_\infty \lesssim 10 \times 10^7$, the generalised forces in transonic buffet can be quite sensitive to the Reynolds number, meaning that the linear scaling may not hold. Given that the phase reduction here is performed at $Re_\infty \approx 3\times10^6$, this study will assess whether the linear scaling can remain useful and informative, and the mechanism by which the phased reduction model departs from the CFD-based aeroelastic result. 

% \textbf{\textit{It is important to emphasise that a substantial reduction has been performed: the high-dimensional transonic buffet fluid state and multi-degree-of-freedom aeroelastic coupling has been reduced to a homogeneous ODE, with a closed-form phase reduced aerodynamic model, and a linear frequency-domain analysis.}} 

\begin{figure}[h!]
	\centering
		\includegraphics[width=1\textwidth]{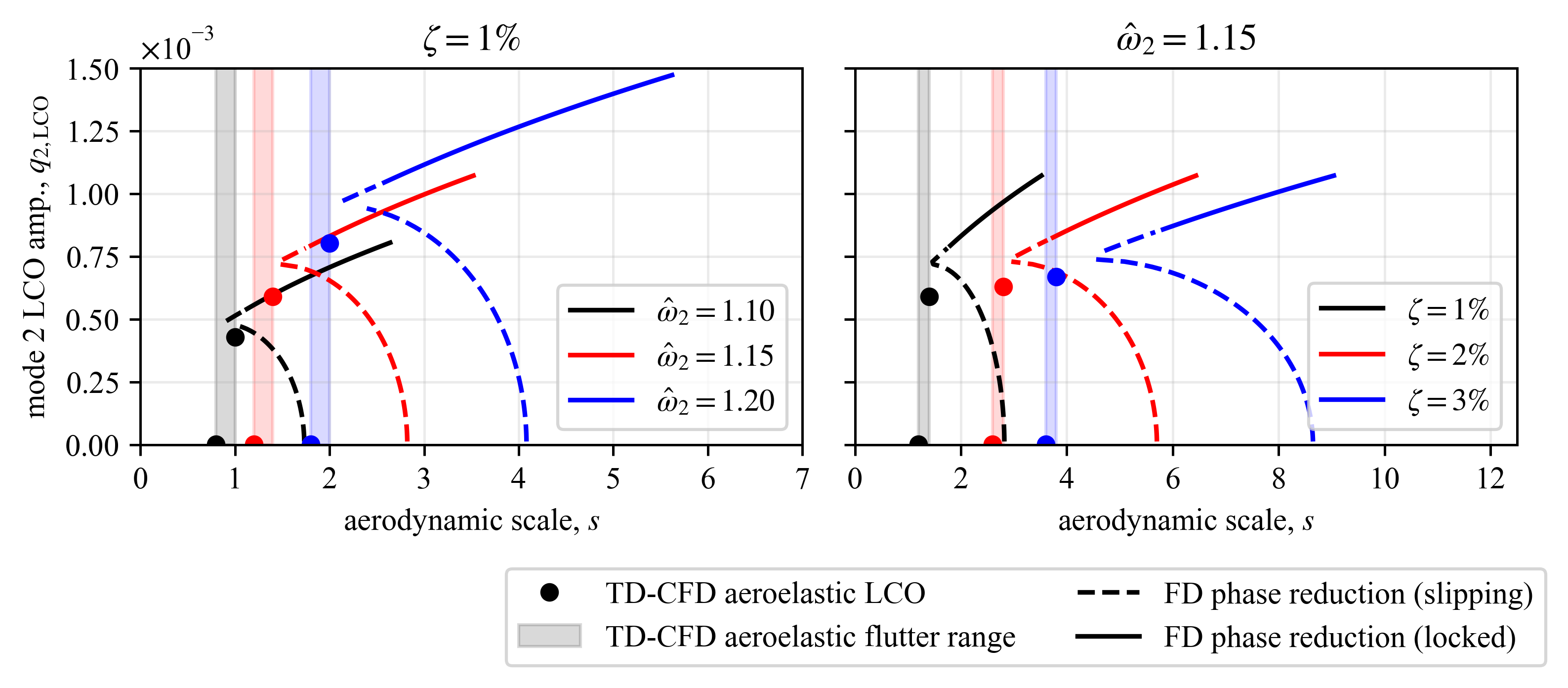}
	\caption{Bifurcation diagrams of the aeroelastic response for $\hat{\omega}_1=0.6$ and $x_\alpha=0.0$.}
	\label{fig:aeroelastic_1}
\end{figure}

Figure~\ref{fig:aeroelastic_1} shows bifurcation diagrams for a structural system that neglects inertial coupling, meaning that the wind-off mode 1 is pure heave and mode 2 is pure pitch. The mode 1 natural frequency is fixed at $\hat \omega_1 = 0.6$, while mode 2 considers $\hat \omega_2 = 1.10, \, 1.15, \, 1.20$. Despite the undamped mode 1 being unstable at this frequency ratio, with structural damping it remains stable for the entire $s$ sweep considered. The structural-to-fluid mass ratio is $\mu = 5000$. Structural damping ratios of $\zeta_1 = \zeta_2 = 1\%, \, 2\%\, 3\%$ are included to achieve a non-zero flutter scale $s_\mathrm{flutter}$. It is incredibly encouraging to see the phase reduced model correctly predicts the single-mode instability of mode 2, despite the substantial reduction of the physics that has taken place. A series of observations from the predictions of the phase reduced aeroelastic models are as follows:

\begin{enumerate}
    \item The unstable branch is approximately predicted by the slipping term of Eqs.~\ref{eq:34} and~\ref{eq:35}, while the stable brach is approximately predicted by the locked term. The accuracy of the predicted subcritical fold is sensitive to the amplitude at which it occurs. 
    \item A significant subritical range exists, with $s_{\mathrm{flutter}}$ occurring at approximately two times the subcritical fold scale, $s_{\mathrm{fold}}$.
    \item From the phase reduced model, $s_{\mathrm{flutter}}$, $s_{\mathrm{fold}}$, and the subcritial fold amplitude $A_\mathrm{fold}$ increase approximately in proportion to the detuning $|\Delta| = |\omega_b - \omega_2|$. 
    \item From the phase reduced model, $s_{\mathrm{flutter}}$ and $s_{\mathrm{fold}}$ increase approximately in proportion to the structural damping ratio $\zeta$. 
    \item From the phase reduced model, $A_\mathrm{fold}$ is largely unaffected by structural damping. 
    \item Relative to the CFD reference, the accuracy of the phase reduction model degrades farther from $s=1$. 
\end{enumerate}

These observations of the phase-reduced aeroelastic dynamics are further explained by examining Eq.~\ref{eq:34}:

\begin{enumerate}
    \item Using $\Omega\approx\omega_2$, the small-amplitude slipping limit of Eq.~\ref{eq:34} gives $|c_a|\propto1/(\omega_2|\Delta|)$ and hence $\boxed{s_{\mathrm{flutter}}\propto\zeta \omega_2^2|\Delta|}$, or $s_{\mathrm{flutter}}\propto\zeta|\Delta|$ over the modest range of $\omega_2$ considered.
    \item Assuming the fold remains close to the transition between the locked and slipping branches, a constant $|\Delta|/(Ak)\approx1$ condition can be applied to Eq.~\ref{eq:34}, and the locked expression gives $\boxed{s_{\mathrm{fold}}\propto\zeta\omega_2^2|\Delta|}$, or $s_{\mathrm{fold}}\propto\zeta|\Delta|$ over the modest range of $\omega_2$ considered.
    \item As $\omega_2$ moves farther from $\omega_B$, the amplitude of the nonlinear fold point increases approximately in proportion to $\Delta$. This comes from the same mechanism as described for $s_\mathrm{fold}$: assuming that the fold occurs at a constant $\Delta/(Ak) \approx 1$, and that $k$ is constant, then $\boxed{A_\mathrm{fold} \propto |\Delta|}$. 
\end{enumerate}

\begin{figure}[h!]
	\centering
		\includegraphics[width=1\textwidth]{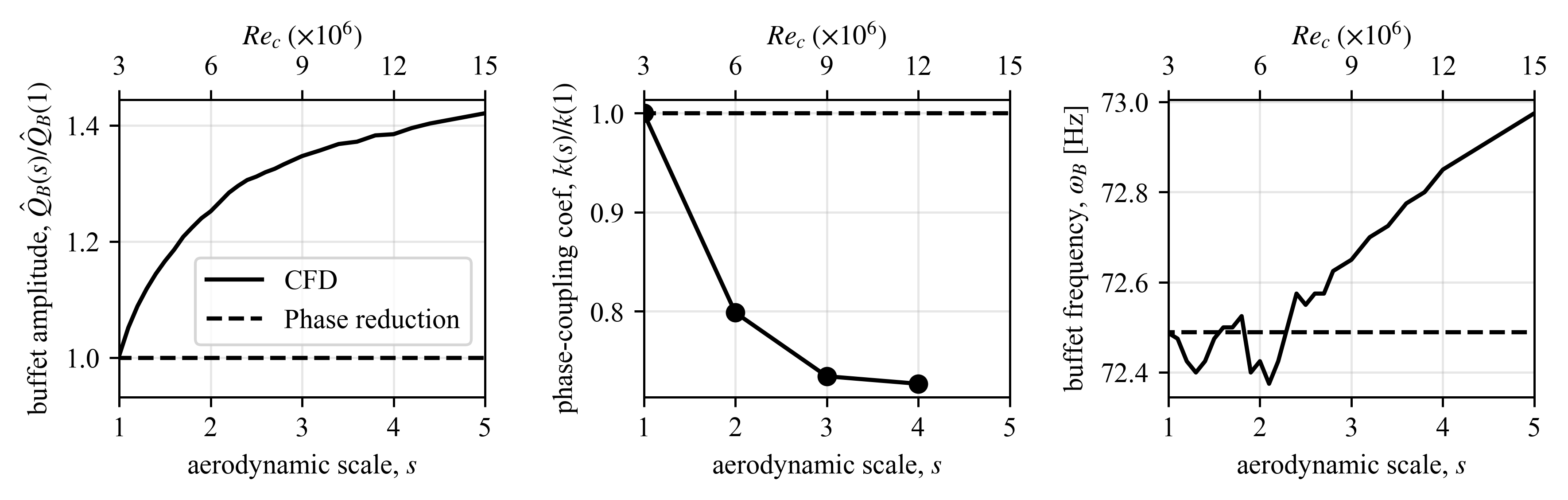}
	\caption{Reynolds number scaling of the buffet amplitude, phase-coupling coefficient, and buffet frequency for the pitch-moment channel.}
	\label{fig:aeroelastic_2}
\end{figure}

The largest assumption of the phase reduction model is that the aerodynamic forces scale linearly with $s$, meaning that $\hat Q_B(s)$, $k(s)$, $\widetilde{\chi}(s)$, and $\omega_B(s)$ are all constants. However, for $\hat Q_B(s)$ and $k(s)$ when $Re_\infty \lesssim 10\times10^6$, this assumption breaks down as can be seen in Fig.~\ref{fig:aeroelastic_2}. Table~\ref{tab:re_scaling} summarises how the different terms in of Eq.~\ref{eq:34} scale for the low- and high-Reynolds regimes. Qualitatively very similar scaling is also shown by Eldridge-Allegra \textit{et al.} ~\cite{eldridgeallegra23}. Taking this scaling into account, for a single-mode instability of a one-degree-of-freedom mode shape (e.g., pure pitch or pure heave), assuming that the fold occurs at $|\Delta|/(Ak)\approx1$, the general scaling of the linear flutter and fold dynamic pressures is:

\begin{equation}
\label{eq:res3}
s_{\mathrm{flutter}}, \, s_{\mathrm{fold}} \propto \frac{\zeta\omega^2|\Delta|}{\hat Q_B k|\sin\widetilde{\chi}|}
\end{equation}

In terms of agreement with the CFD-based aeroelastic predictions, considering the substantial reduction of the physics that has taken place, and that the verification is being performed at the nonlinear flutter point (which the model is not expected to predict well anyway), the results are very good and undoubtedly informative. The minimum error in the prediction of $s_\mathrm{fold}$ and $A_\mathrm{fold}$ is $~\sim 5\%$ and the maximum is $~\sim 30\%$. 

The issue now is that Eq.~\ref{eq:res3} does not account for the degrading accuracy of the phase reduction aeroelastic model for much larger values of $s$. This is because the increase in $\hat Q_B$ and decrease in $k$ cancel each other, and therefore Eq.~\ref{eq:res3} cannot be responsible for this observation. Note that this cancelling should not be considered a general rule, but certainly does apply here. Therefore, an explanation for this comes from the static aeroelastic deformation not being taken into account which can have a non-negligible influence $\omega_B$. More specifically, a small change in the buffet frequency due to a static offset of $0.1^\circ$ or $0.2^\circ$ can have a substantial influence on the dynamics when the detuning (as calculated using the rigid aerofoil buffet frequency, $\omega_B$) is very close to 1. This hypotheses is left open and deserves further in future work. If true, this effect should become less noticeable as it is absorbed by larger $\Delta \omega$ values, or for stiffer models. Another way to test this would be through single-degree-of-freedom heave instabilities where this effect should not exist at all.

\begin{figure}[h!]
	\centering
		\includegraphics[width=0.625\textwidth]{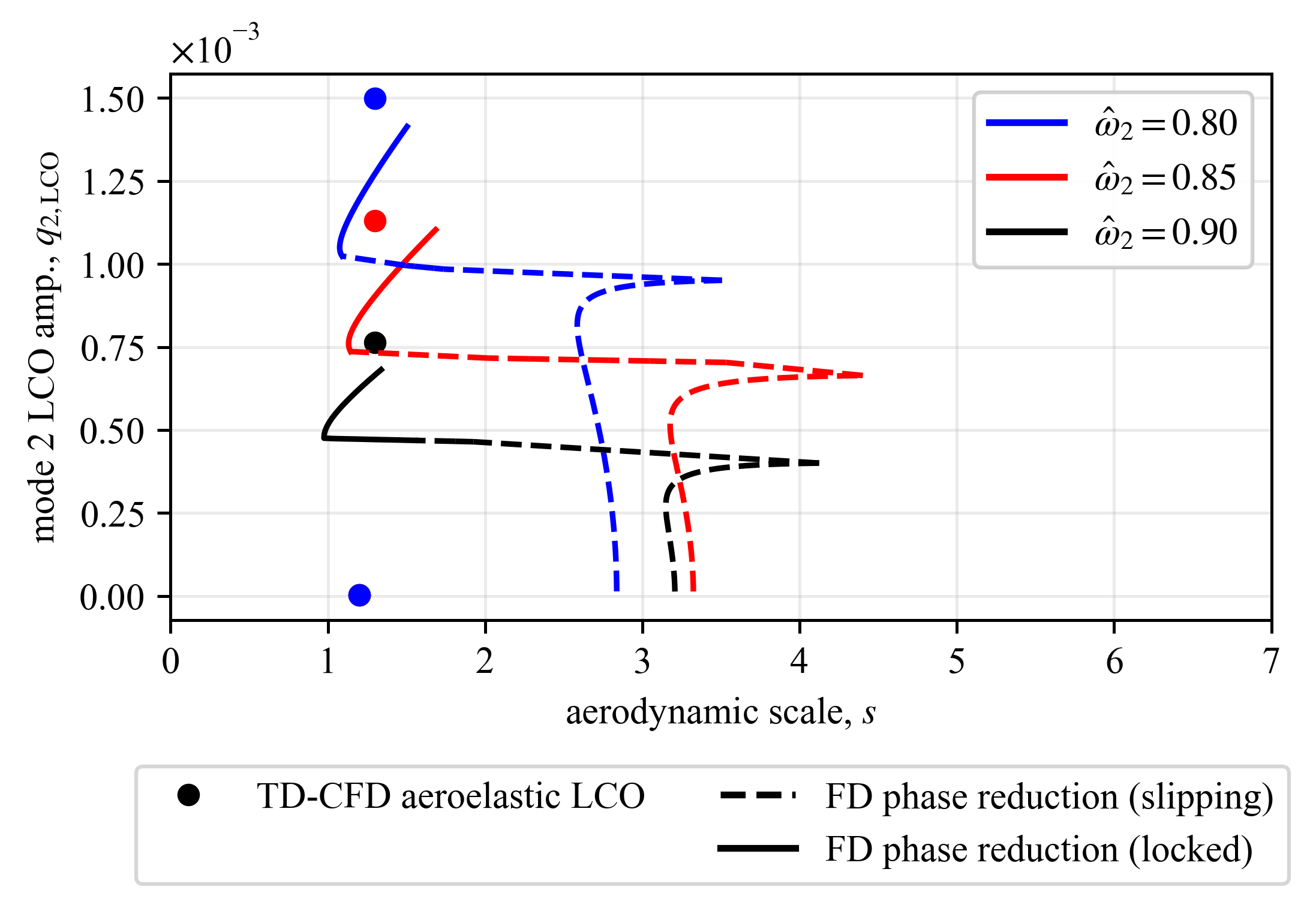}
	\caption{Bifurcation diagrams of the aeroelastic response for $\hat{\omega}_1=0.4$ and $x_\alpha=0.5$.}
	\label{fig:aeroelastic_3}
\end{figure}

\begin{table}[!h]
	\centering
	\caption{Scaling behaviour of the buffet and phase-reduction quantities with Reynolds number.}
    \label{tab:re_scaling}
	\begin{tabular}{lccc}
		\hline
		Quantity & Phase Reduction & $Re_\infty > 10M$ & $Re_\infty < 10M$ \\
		\hline
		$\hat{Q}_B$ & $\mathrm{const.}$ & $\sim \mathrm{const.}$ & $f(Re_\infty)$ \\
		$k$ & $\mathrm{const.}$ & $\sim \mathrm{const.}$ & $f(Re_\infty)$ \\
		$\sin(\widetilde{\chi})$ & $\mathrm{const.}$ & $\sim \mathrm{const.}$ & $\approx \mathrm{const.}$ \\
		$\omega_B$ & $\mathrm{const.}$ & $\sim \mathrm{const.}$ & $\sim \mathrm{const.}$ \\
		\hline\hline
	\end{tabular}
\end{table}
The second aeroelastic case, presented in Fig.~\ref{fig:aeroelastic_3}, includes substantial inertial coupling, such that the wind-off mode 1 is heave dominated but contains a pitch component, whereas mode 2 is pitch dominated but contains a heave component. In this case, no clear trend is observed in either $s_{\mathrm{flutter}}(\omega_2)$ or $s_{\mathrm{fold}}(\omega_2)$, both of which remain nearly constant over the range considered. Referring back to Fig.~\ref{fig:aerodamping_B}, as $\Omega/\omega_B$ decreases below unity, the negative heave damping becomes less negative while the positive pitch damping becomes less positive. These competing contributions therefore offset one another in the coupled heave-pitch mode shape. The relationship $A_{\mathrm{fold}}\propto|\Delta|$ nevertheless remains. The other primary feature is the substantially deeper subcritical region, with the linear flutter point occurring at approximately three times the aerodynamic scale of the nonlinear fold.

\begin{figure}[h!]
	\centering
		\includegraphics[width=0.5625\textwidth]{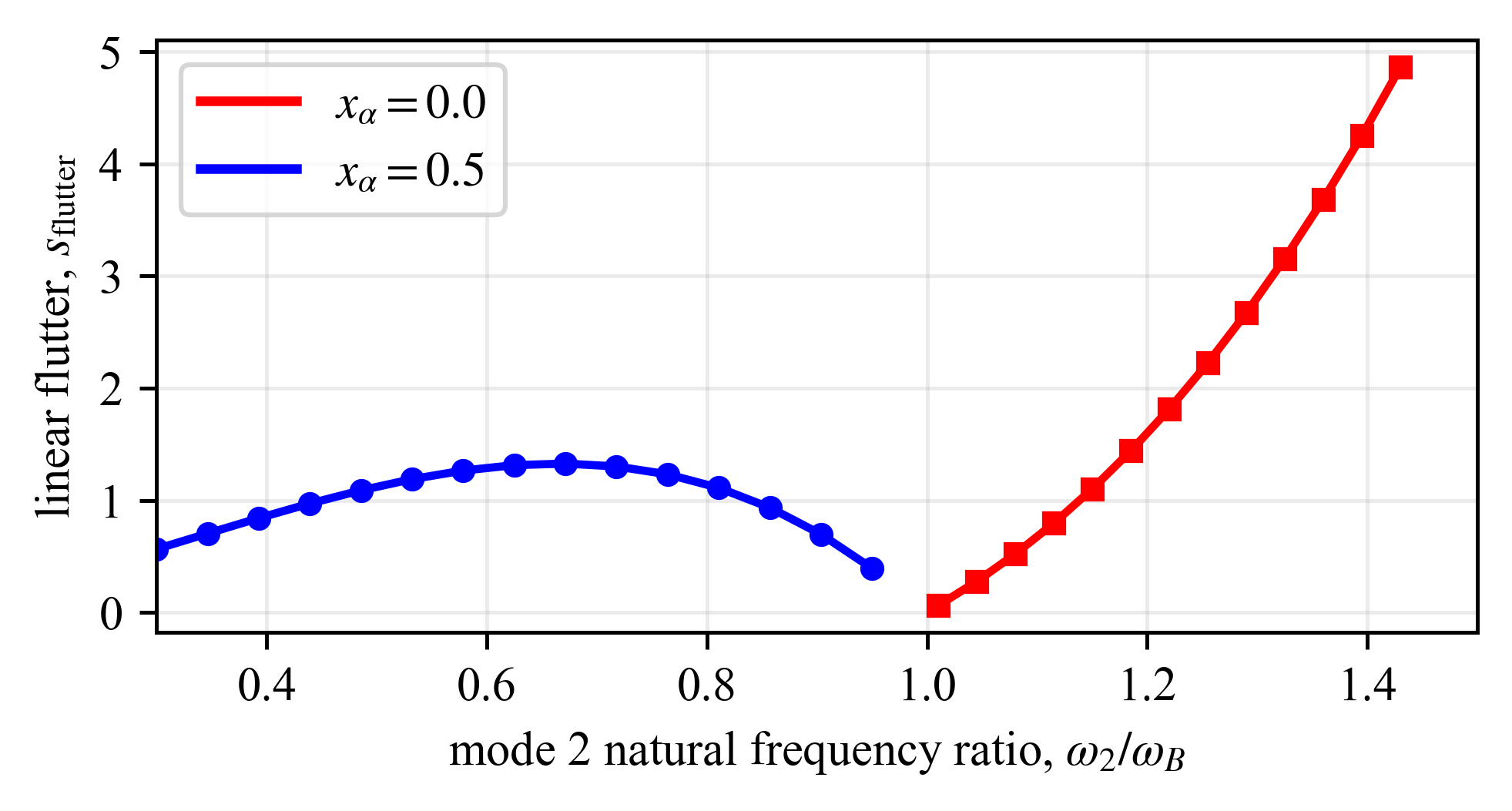}
	\caption{Linear flutter speed with fixed ratios of the mode 1-to-2 natural frequency.}
	\label{fig:aeroelastic_4}
\end{figure}

This contrast is clearer in Fig.~\ref{fig:aeroelastic_4}, where $\omega_1/\omega_2=0.51$ is held fixed while $\omega_2$ is varied. For the uncoupled mode shape, $s_{\mathrm{flutter}}$ increases approximately as $\omega_2|\Delta|$, so moving the natural frequencies away from the buffet frequency substantially increases the stability margin. For the coupled mode shape, however, $s_{\mathrm{flutter}}$ remains of $\mathcal{O}(1)$ across the range considered. Inertial coupling therefore removes the expected benefit of frequency separation: moving the structural frequencies farther from the buffet frequency does not materially delay the instability.

\section{Conclusions}

Phase reduction has been applied to transonic buffet to establish a direct connection between synchronisation dynamics and conventional aeroelastic stability. Closed-form expressions for the equivalent aerodynamic damping and stiffness were derived, describing both phase-locked and phase-slipping dynamics and revealing a singular aerodynamic response at zero detuning. The resulting model captures the dominant aerodynamic damping characteristics and predicts broad regions of subcritical aeroelastic instability. Despite the substantial reduction of the underlying fluid dynamics, the frequency-domain aeroelastic predictions show encouraging agreement with time-domain CFD, demonstrating that phase reduction provides a useful framework for understanding and predicting buffet-induced aeroelastic instability.

\appendix

\section{Numerical Convergence and Experimental Validation}
\label{appA}
 
\begin{figure}[!h]
    \centering
    \subfigure[Combined spatial and temporal refinement of the pressure statistics]{%
    \label{fig:gridref}
    \includegraphics[width=0.95\textwidth]{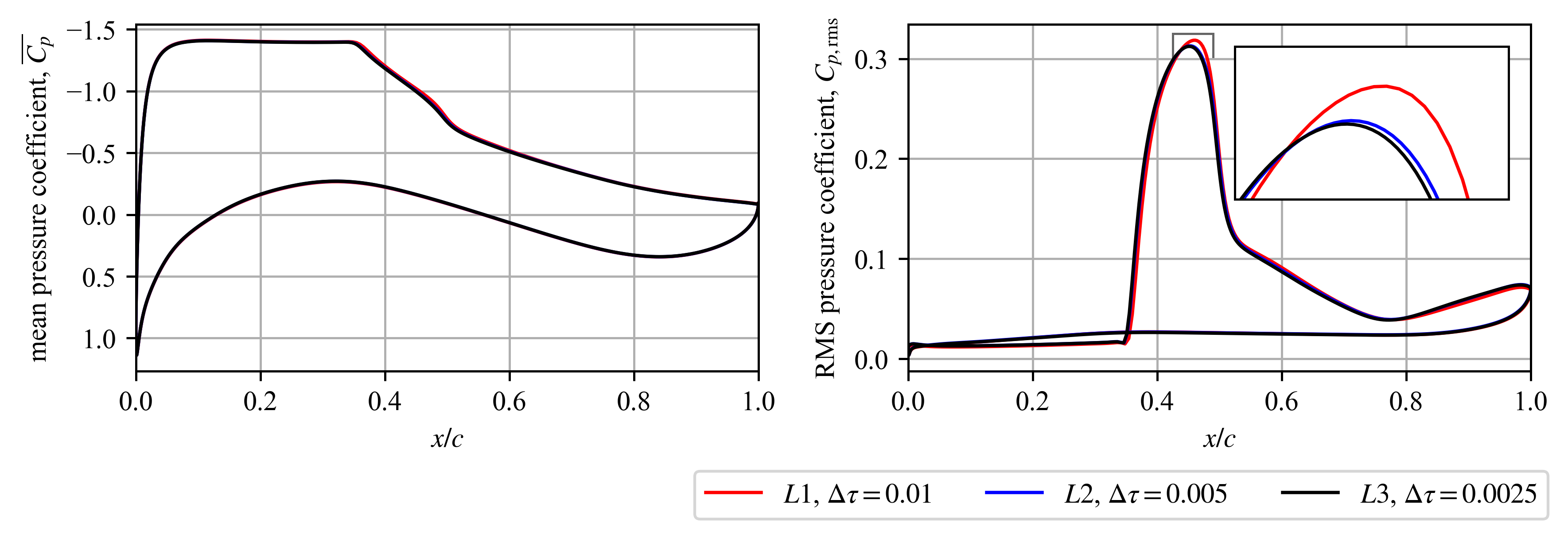}}

    \subfigure[Comparison of the computed and experimental pressure statistics]{%
    \label{fig:expval}
    \includegraphics[width=0.95\textwidth]{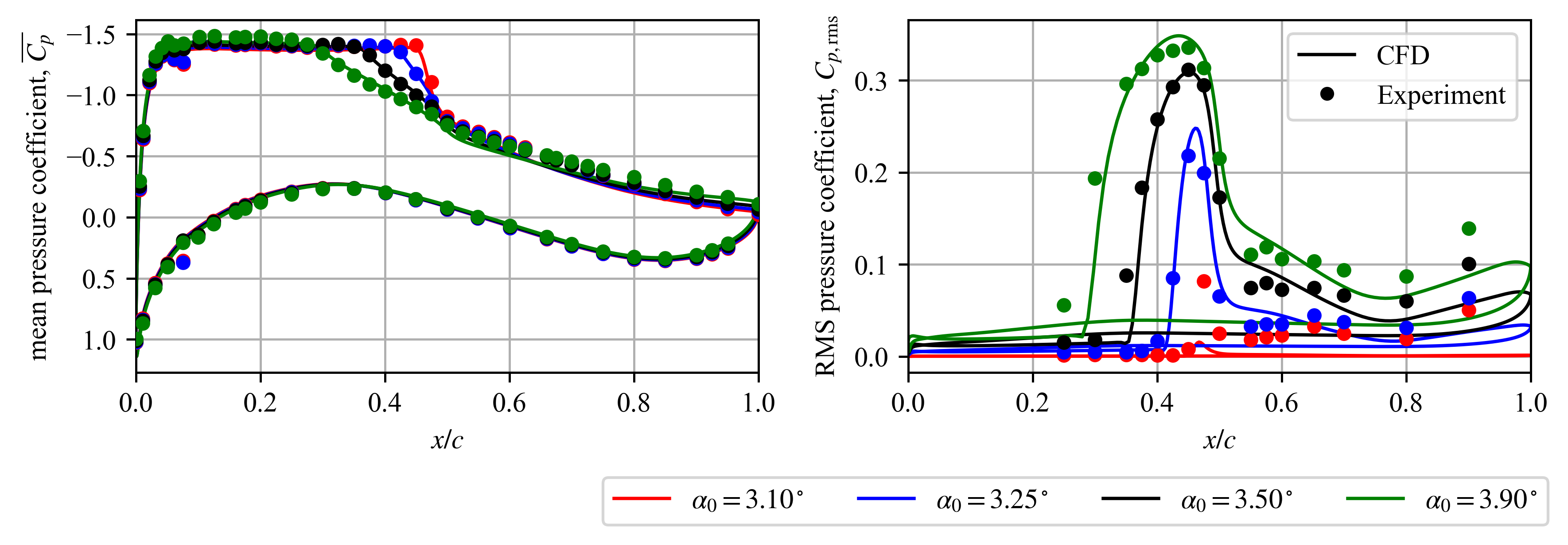}}

    \caption{Numerical refinement and experimental validation.}
    \label{fig:validation}
\end{figure}

A combined spatial and temporal refinement study~\cite{housman11} is shown in Fig.~\ref{fig:validation}(a) and summarised in Table~\ref{tab:mesh}. For each grid level, the characteristic cell size in the shock region is approximately halved and the non-dimensional time step is correspondingly reduced by a factor of two, thereby maintaining a comparable local CFL. From L2--$\Delta\tau=0.005$ to L3--$\Delta\tau=0.0025$, the changes in the peak-to-peak lift and pitching-moment coefficients are both less than 2\%, while the buffet frequency is effectively unchanged. The L2 mesh with $\Delta\tau=0.005$ was therefore selected for the remaining calculations as a conservative compromise between computational cost and numerical accuracy.

The GEKO model is calibrated against the experimental data at $\alpha_0=3.50^\circ$ by adjusting the separation parameter $C_{\mathrm{SEP}}$, which modifies the eddy viscosity and hence the adverse-pressure-gradient separation behaviour without compromising the logarithmic-layer calibration~\cite{menter25}. As shown in Fig.~\ref{fig:validation}(b), the resulting model reproduces the measured mean and RMS pressure distributions at the other angles of attack with reasonable fidelity. This agreement outside the calibration condition indicates that the model is sufficiently general and has not been overfit to the $\alpha_0=3.50^\circ$ case.

\begin{table}[!ht]
	\centering
	\caption{Computational grid and time-step statistics and convergence of the integrated aerodynamic forces. Values in parentheses are the absolute percentage change relative to the preceding case.}
    \label{tab:mesh}
	\begin{tabular}{cccccccc}
		\hline
		Mesh & Cells & Growth rate & $y^+$ & $\Delta \tau$ & $\Delta C_L$ [\%] & $\Delta C_M$ [\%] & $\omega_B$ [Hz] [\%] \\
		\hline
		L1 & 51,928  & 1.16 & $\sim 1.00$ & 0.0100 & 0.196 (--) & 0.052 (--) & 73.10 (--) \\
		L2 & 118,300 & 1.10 & $\sim 0.67$ & 0.0050 & 0.203 (3.18\%) & 0.052 (1.59\%) & 73.04 (0.09\%) \\
		L3 & 232,935 & 1.07 & $\sim 0.50$ & 0.0025 & 0.200 (1.00\%) & 0.051 (1.80\%) & 73.03 (0.00\%) \\
		\hline\hline
	\end{tabular}
\end{table}

\section{Flutter Sensitivity to Phase Reduction}
\label{appB}

\begin{figure}
	\centering
		\includegraphics[width=1\textwidth]{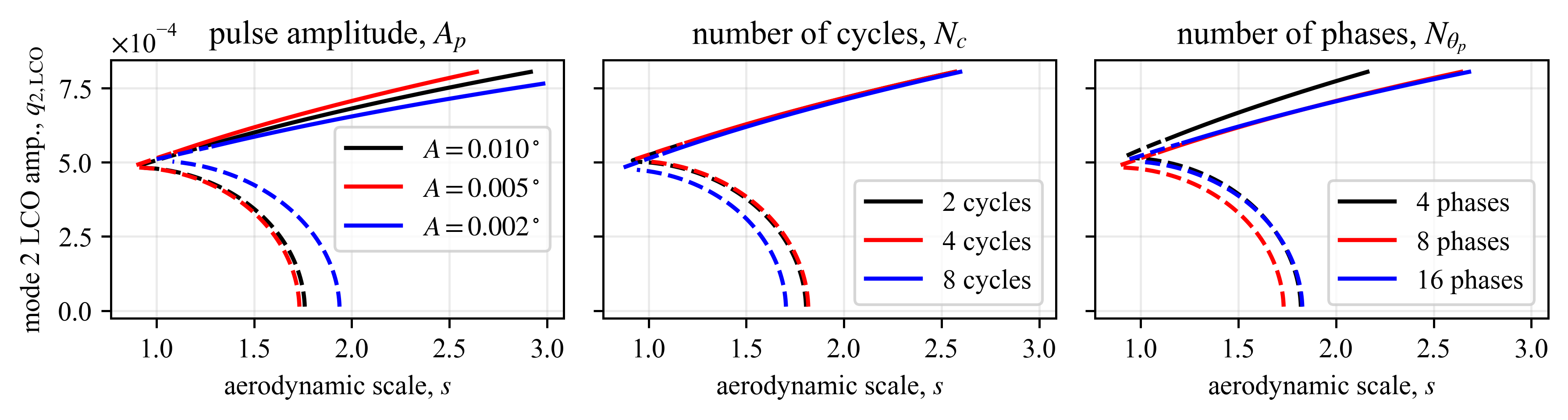}
	\caption{Sensitivity of the aeroelastic bifurcation diagrams to phase reduction parameters.}
	\label{fig:aeroelastic_sens}
\end{figure}

From a practical perspective, it is important to understand how sensitive the phase reduction is to a number of parameters. The parameters presented in Fig.~\ref{fig:aeroelastic_sens} include the pulse amplitude, $A_p$, the number of cycles retained after the pulse is applied, $N_c$, and the resolution of the phase discretisation of the buffet limit cycle, $N_{\theta_p}$. Before discussing the results, the reader should recall that from an applied perspective, this method is not intended to give flutter predictions to within a few percent of the full-order CFD model, but rather to rapidly identify if structural modes are unstable under buffet and give an indication of their flutter speed (to within $\sim20\%$) and character. This is information that presently would be out of reach when considering the design of aircraft. The most important finding is that the flutter result is reasonably insensitive to both $N_c$ and $N_{\theta_p}$ which means that with as few as four evenly spaced pulses of the buffet cycle ($0, \, \pi/2, \, \pi, \, 3\pi/2$) and four buffet cycles post pulse (discarding the first two and measuring the phase sensitivity from the second two). 

\section{Acknowledgments}
We are grateful for the financial support of the Asian Office of Aerospace Research and Development (AOARD) and Air Force Office of Scientific Research (AFOSR) for project FA2386-24-1-4044: Data-Driven Reduced Order Modelling and Preliminary Experimentation for Combined Transonic Buffet and Freeplay Induced Limit Cycle Oscillations. The generous resources and support of Dale Osborne and Dr. Robert Shen from RMIT RACE Cloud Supercomputing Hub, and the resources provided by ANSYS and the support of Dr Valerio Viti and Dr Luke Munholand, are also all greatly appreciated.

% Bibliography
% ------------
% \bibliographystyle{IEEEtranDOI}
\bibliographystyle{elsarticle-num} 
\bibliography{gensys_doi}
        
%         \bibliography{gensys_doi_2}

\end{document}